\documentclass[nolinenumbers, twocolumn]{aastex631}

\def\tess{$\it{TESS}$}

\newcommand{\Mpl}{$14.5_{-3.14}^{+3.15} ~\rm{M_\oplus}$}
\newcommand{\Rpl}{$2.77_{-0.07}^{+0.15}~\rm{R_\oplus}$}
\newcommand{\rhopl}{$3.6 \pm 0.9 \, {\rm g}\,{\rm cm}^{-3}$}

\graphicspath{{Figures/}{}}

\begin{document}
\title{The TESS-Keck Survey. XVIII. \\ A sub-Neptune and spurious long-period signal in the TOI-1751 system}
\author[0009-0000-9206-5589]{Anmol Desai}
\affiliation{Department of Astronomy, 501 Campbell Hall, University of California, Berkeley, CA 94720, USA}

\author[0000-0002-1845-2617]{Emma V. Turtelboom}
\affiliation{Department of Astronomy, 501 Campbell Hall, University of California, Berkeley, CA 94720, USA}

\author[0000-0001-5737-1687]{Caleb K. Harada}
\altaffiliation{NSF Graduate Research Fellow}
\affiliation{Department of Astronomy, 501 Campbell Hall, University of California, Berkeley, CA 94720, USA}

\author[0000-0001-8189-0233]{Courtney D. Dressing}
\affiliation{Department of Astronomy, 501 Campbell Hall, University of California, Berkeley, CA 94720, USA}

\author[0000-0001-6009-8685]{David R. Rice}
\affiliation{Astrophysics Research Center, Department of Natural Sciences, The Open University of Israel, Raanana 4353701, Israel}

\author[0000-0001-8898-8284]{Joseph M. Akana Murphy}
\altaffiliation{NSF Graduate Research Fellow}
\affiliation{Department of Astronomy and Astrophysics, University of California, Santa Cruz, CA 95064, USA}

\author[0000-0002-4480-310X]{Casey L. Brinkman}
\affiliation{Institute for Astronomy, University of Hawai`i, 2680 Woodlawn Drive, Honolulu, HI 96822, USA}

\author[0000-0003-1125-2564]{Ashley Chontos}
\altaffiliation{Henry Norris Russell Fellow}
\affiliation{Department of Astrophysical Sciences, Princeton University, 4 Ivy Lane, Princeton, NJ 08540, USA}

\author{Ian J.\ M.\ Crossfield}
\affiliation{Department of Physics and Astronomy, University of Kansas, Lawrence, KS, USA}

\author[0000-0002-8958-0683]{Fei Dai}
\altaffiliation{NASA Sagan Fellow}
\affiliation{Division of Geological and Planetary Sciences,
1200 E California Blvd, Pasadena, CA, 91125, USA}
\affiliation{Department of Astronomy, California Institute of Technology, Pasadena, CA 91125, USA} 

\author[0000-0002-0139-4756]{Michelle L. Hill}
\affiliation{Department of Earth and Planetary Sciences, University of California, Riverside, CA 92521, USA}

\author[0000-0002-3551-279X]{Tara Fetherolf}
\affiliation{Department of Physics, California State University, San Marcos, CA 92096, USA}
\affiliation{Department of Earth and Planetary Sciences, University of California, Riverside, CA 92521, USA}

\author[0000-0002-8965-3969]{Steven Giacalone}
\altaffiliation{NSF Astronomy and Astrophysics Postdoctoral Fellow}
\affiliation{Department of Astronomy, California Institute of Technology, Pasadena, CA 91125, USA}

\author[0000-0001-8638-0320]{Andrew W.\ Howard}
\affiliation{Cahill Center for Astronomy $\&$ Astrophysics, California Institute of Technology, Pasadena, CA 91125, USA}

\author[0000-0001-8832-4488]{Daniel Huber}
\affiliation{Institute for Astronomy, University of Hawai`i, 2680 Woodlawn Drive, Honolulu, HI 96822, USA}
\affiliation{Sydney Institute for Astronomy (SIfA), School of Physics, University of Sydney, NSW 2006, Australia}

\author[0000-0002-0531-1073]{Howard Isaacson}
\affiliation{Department of Astronomy, 501 Campbell Hall, University of California, Berkeley, CA 94720, USA}
\affiliation{Centre for Astrophysics, University of Southern Queensland, Toowoomba, QLD, Australia}

\author[0000-0002-7084-0529]{Stephen R. Kane}
\affiliation{Department of Earth and Planetary Sciences, University of California, Riverside, CA 92521, USA}

\author[0000-0001-8342-7736]{Jack Lubin}
\affiliation{Department of Physics \& Astronomy, University of California Irvine, Irvine, CA 92697, USA}

\author[0000-0003-2562-9043]{Mason G.\ MacDougall}
\affiliation{Department of Physics \& Astronomy, University of California Los Angeles, Los Angeles, CA 90095, USA}

\author[0000-0002-7216-2135]{Andrew W. Mayo}
\affiliation{Department of Astronomy, 501 Campbell Hall, University of California, Berkeley, CA 94720, USA}
\affiliation{Centre for Star and Planet Formation, Natural History Museum of Denmark \& Niels Bohr Institute, University of Copenhagen, \O ster Voldgade 5-7, DK-1350 Copenhagen K., Denmark}

\author[0000-0003-4603-556X]{Teo Mo\v{c}nik}
\affiliation{Gemini Observatory/NSF's NOIRLab, 670 N. A'ohoku Place, Hilo, HI 96720, USA}

\author[0000-0001-7047-8681]{Alex S. Polanski}
\affiliation{Department of Physics and Astronomy, University of Kansas, Lawrence, KS, USA}

\author[0000-0002-7670-670X]{Malena Rice}
\affiliation{Department of Astronomy, Yale University, New Haven, CT 06511, USA}

\author[0000-0003-0149-9678]{Paul Robertson}
\affiliation{Department of Physics \& Astronomy, University of California Irvine, Irvine, CA 92697, USA}

\author[0000-0003-3856-3143]{Ryan A. Rubenzahl}
\altaffiliation{NSF Graduate Research Fellow}
\affil{Department of Astronomy, California Institute of Technology, Pasadena, CA 91125, USA}

\author[0000-0002-4290-6826]{Judah Van Zandt}
\affiliation{Department of Physics \& Astronomy, University of California Los Angeles, Los Angeles, CA 90095, USA}

\author[0000-0002-3725-3058]{Lauren M. Weiss}
\affil{Department of Physics and Astronomy, University of Notre Dame, Notre Dame, IN 46556, USA}

\author[0000-0001-6637-5401]{Allyson~Bieryla} 
\affiliation{Center for Astrophysics ${\rm \mid}$ Harvard {\rm \&} Smithsonian, 60 Garden Street, Cambridge, MA 02138, USA}

\author[0000-0003-1605-5666]{Lars A. Buchhave}
\affiliation{DTU Space,  Technical University of Denmark, Elektrovej 328, DK-2800 Kgs. Lyngby, Denmark}

\author[0000-0002-4715-9460]{Jon~M.~Jenkins}
\affiliation{NASA Ames Research Center, Moffett Field, CA 94035, USA}

\author[0000-0001-9786-1031]{Veselin~B.~Kostov}
\affiliation{NASA Goddard Space Flight Center, 8800 Greenbelt Road, Greenbelt, MD 20771, USA}
\affiliation{SETI Institute, 189 Bernardo Ave, Suite 200, Mountain View, CA 94043, USA}

\author[0000-0001-8172-0453]{Alan M. Levine}
\affiliation{Department of Physics and Kavli Institute for Astrophysics and Space Research, Massachusetts Institute of Technology, Cambridge, MA 02139, USA}

\author[0000-0003-3742-1987]{Jorge Lillo-Box}
\affiliation{Centro de Astrobiolog\'ia (CAB), CSIC-INTA, Depto. de Astrof\'isica, ESAC campus, 28692, Villanueva de la Ca\~nada (Madrid), Spain}

\author[0000-0001-8120-7457]{M. Paegert}
\affiliation{Center for Astrophysics ${\rm \mid}$ Harvard {\rm \&} Smithsonian, 60 Garden Street, Cambridge, MA 02138, USA}

\author[0000-0003-2935-7196]{Markus Rabus} 
\affiliation{Departamento de Matem{\'a}tica y F{\'i}sica Aplicadas, Facultad de Ingenier{\'i}a, Universidad Cat{\'o}lica de la Sant{\'i}sima Concepci{\'o}n, Alonso de Rivera 2850, Concepci{\'o}n, Chile }

\author[0000-0002-6892-6948]{S.~Seager}
\affiliation{Department of Physics and Kavli Institute for Astrophysics and Space Research, Massachusetts Institute of Technology, Cambridge, MA 02139, USA}
\affiliation{Department of Earth, Atmospheric and Planetary Sciences, Massachusetts Institute of Technology, Cambridge, MA 02139, USA}
\affiliation{Department of Aeronautics and Astronautics, MIT, 77 Massachusetts Avenue, Cambridge, MA 02139, USA}

\author[0000-0002-3481-9052]{Keivan G.\ Stassun}
\affiliation{Department of Physics and Astronomy, Vanderbilt University, Nashville, TN 37235, USA}

\author[0000-0002-8219-9505]{Eric B. Ting}
\affiliation{NASA Ames Research Center, Moffett Field, CA 94035, USA}

\author[0000-0002-3555-8464]{David Watanabe}
\affiliation{Planetary Discoveries, 28935 Via Adelena, Santa Clarita CA 91354}

\author[0000-0002-4265-047X]{Joshua N.\ Winn}
\affiliation{Department of Astrophysical Sciences, Princeton University, Princeton, NJ 08544, USA}

\correspondingauthor{Emma Turtelboom}
\email{eturtelboom@berkeley.edu}

\begin{abstract}
We present and confirm TOI-1751~b, a transiting sub-Neptune orbiting a slightly evolved, solar-type, metal-poor star ($T_{eff} = 5996 \pm 110$ K, $log(g) = 4.2 \pm 0.1$, V = 9.3 mag, [Fe/H] = $-0.40 \pm 0.06 dex$) every 37.47~d. We use \tess{} photometry to measure a planet radius of \Rpl{}. We also use both Keck/HIRES and APF/Levy radial velocities (RV) to derive a planet mass of \Mpl{}, and thus a planet density of \rhopl{}. There is also a long-period ($\sim400~\rm{d}$) signal that is observed in only the Keck/HIRES data. We conclude that this long-period signal is not planetary in nature, and is likely due to the window function of the Keck/HIRES observations. This highlights the role of complementary observations from multiple observatories to identify and exclude aliases in RV data. Finally, we investigate potential compositions of this planet, including rocky and water-rich solutions, as well as theoretical irradiated ocean models. TOI-1751~b is a warm sub-Neptune, with an equilibrium temperature of $\sim 820$ K.  As TOI-1751 is a metal-poor star, TOI-1751~b may have formed in a water-enriched formation environment. We thus favor a volatile-rich interior composition for this planet.

\end{abstract}
\section{Introduction} \label{sec:intro}
The Transiting Exoplanet Survey Satellite (\tess{}, \citealt{tess}) has discovered 415 confirmed exoplanets to date\footnote{\label{note1}NASA Exoplanet Archive, \url{https://exoplanetarchive.ipac.caltech.edu/}, accessed 31 January 2024}, and has identified thousands of planet candidates. These planets join the extensive population of over 5500 known planets. With this sample, we can search for demographic trends in planet radii, masses, compositions, and occurrence, and begin to probe the mechanisms of planet formation and evolution that govern the planets in our galaxy.

The most commonly detected exoplanets are those smaller than Neptune ($1-4~\rm{R_\oplus}$, commonly referred to as sub-Neptunes and/or super-Earths${}^1$). \tess{} has discovered over 200 such planets, and $\gtrsim75\%$ of all confirmed planets with measured radii are smaller than 4 $\rm{R_\oplus} {}^1$. Population-level trends offer windows into these planets' evolution and formation.

One such feature is the radius gap, a valley in the distribution of planet radii near $1.8~\rm{R_\oplus}$ \citep{fulton+17}. This feature may be explained by two distinct populations: primarily rocky planets with H/He envelopes, and planets with volatile-rich interiors and more voluminous envelopes \citep[e.g.][]{zeng+2019, izidoro+22, venturini+17, venturini+20}. \citet{eve+22} suggest the radius gap is imprinted on the planet population at early times through planet formation in gas-poor disks. Alternatively, \citet{luque+22} posit that the two populations are distinct in density, rather than radius, around M dwarfs. In this framework, compositional differences between the populations are set by the materials available in their formation environments, suggesting distinct formation locations within their protoplanetary disks. Water-rich planets may form beyond the snow line and then migrate inwards \citep{legre+04, bitsch+19}. On the other hand, rocky planets may form in situ from ice-poor pebbles, and accrete primordial atmospheres from their natal disks \citep{lee+16, chiang+13}. 

Other mechanisms involving the sculpting of a single underlying population to create the radius gap have also been suggested. These include atmospheric mass-loss through XUV-driven photoevaporation \citep{owen+wu13, rogers+21}, core-powered mass loss \citep{ginzburg+18}, and atmospheric mass-loss and growth through giant impacts \citep{wyatt+20}. 

In this paper, we present and confirm TOI-1751~b, a sub-Neptune with a period of 37.47 days, orbiting a slightly evolved G0 star. TOI-1751~b is at the upper radius boundary of the sub-Neptune population, where planet occurrence begins to decrease. While much work has been done to investigate the mechanism(s) sculpting the radius gap, there are few models to explain this ``occurrence cliff'' \citep{dattilo+23}. By characterizing planets in this regime, we may refine models that encapsulate both the radius gap and occurrence cliff. Furthermore, due to the lower transit probability for planets with longer periods, only about $10\%$ of sub-Neptunes confirmed by \tess{} have orbital periods greater than 25 d. TOI-1751~b thus resides in a population that is challenging to probe with transit surveys.

This target was observed by the \tess-Keck Survey (TKS), a collaboration spanning several institutions that pools time on the Keck-I telescope on Maunakea. It was initially selected for radial velocity (RV) observations under several science cases: searching for distant giants, probing planets across the radius gap, and analyzing the diversity of gaseous envelopes (see \citealt{tks0} for a comprehensive of the TKS science cases). In Section \ref{sec:data} we discuss the data collected for this target. In Section \ref{sec:star} we characterize the stellar host. Section \ref{sec:analysis} presents our photometric and RV analyses and resulting planet parameters. In Section \ref{sec:discussion} we put this system in context and discuss possible planetary compositions, and conclude in Section \ref{sec:conc}.

\section{Data Collected} \label{sec:data}
\subsection{Photometric Observations} \label{sec:tess}
The \tess{} mission observed TOI-1751 (TIC 287080092, HD 146757) for a total of 27 sectors (15, 17-26, 40-41, and 47-59) between 15 August 2019 and 23 December 2022 at 120 s cadence. The Science Processing Operations Center (SPOC, \citealt{jenkinsSPOC2016}) conducted a transit search of Sectors 15, 17, 18 and 19 on 24 January 2020 with an adaptive, noise-compensating matched filter \citep{2002ApJ...575..493J, 2010SPIE.7740E..0DJ, 2020TPSkdph}. This produced a threshold-crossing event (TCE) to which an initial limb-darkened transit model was fit \citep{Li:DVmodelFit2019}. Diagnostic tests were then conducted to investigate the planetary nature of the signal \citep{Twicken:DVdiagnostics2018}. The transit signal passed all of the diagnostic tests. The TESS Science Office (TSO) reviewed the vetting information and issued an alert on 27 February 2020 \citep{guerrero:TOIs2021ApJS}. The signal was repeatedly recovered as additional observations were made in sectors 20-26, 40-41 and 47-59. The final transit search located the host star within $5.8 \pm 3.6 \arcsec$ of the source of the transit signal using a difference image centroiding test. We note that observations taken in sectors 15 and 17-26 were impacted by a bias in the sky background correction algorithm, which tended to overestimate the sky background flux. However, the impact on the derived planetary radius of TOI-1751~b is below $0.4\%$ in all affected sectors and typically between $0.1\%$ – $0.2\%$, so this is not a dominant error source in our analysis.

The target was also observed at 1800-s cadence in sector 16 and processed from Full-Frame Images (FFIs) through the Quick Look Pipeline \citep{2020RNAAS...4..204H, 2020RNAAS...4..206H}. During sectors 56-59, TOI-1751 was also observed at 20-s cadence. This target was part of the following \tess{} Guest Observer programs: G04242 (PI: Mayo), G04191 (PI: Burt), G04039 (PI: Davenport), and G05144 (20-second target, PI: Huber).

\subsection{Imaging Observations}
Eclipsing binaries with small projected separations from a putative planet host star can create a false-positive transit signal. This effect is particularly important to consider for \tess{} Objects of Interest (TOIs), due to the \tess{} mission's larger pixel size ($21\arcsec \times 21\arcsec$) compared to that of \textit{Kepler} ($4\arcsec \times 4\arcsec$). The SPOC transit search of TOI-1751 was able to constrain the location of the transit event to within a \tess{} pixel (see Section \ref{sec:tess}), suggesting a blended eclipsing binary was not the cause of the transit events. However, the flux from companion stars can lead to underestimated planetary radii, overestimated bulk densities, and erroneous stellar parameters  \citep{ciardi+2015, furlan+2017}. The contaminating flux from a nearby star may also inhibit the detection of shallow transits \citep{lester+2021}. We thus obtained high-resolution imaging observations of TOI-1751 to search for nearby companion stars. 
 
\subsubsection{Lucky Imaging}
We obtained two observations of TOI-1751 with the AstraLux instrument (\citealt{hormuth08}) installed at the 2.2-m telescope in the Calar Alto Observatory (Almer\'{i}a, Spain) under average weather and atmospheric conditions (seeing around 1\arcsec) on 22 March and 14 September 2021 with the SDSS z filter. AstraLux uses the lucky imaging technique to obtain thousands of short exposure frames and selects a few percent of these frames with the best Strehl ratio \citep{strehl1902}. We obtained 62,200 frames with an exposure time of \mbox{10 ms} each and selected the best 10\% for a final effective exposure time of 62.2\,s. We used the final stacked image to obtain the contrast curve by using the \texttt{astrasens} code \citep{lillobox+2012,lillo-box14b}. The result provides a contrast of $\Delta z=6.3$~mag for separations above 0.5\arcsec on the 22 March 2021. We found no additional sources in the field of view of the instrument ($3\arcsec\times 3\arcsec$) within these sensitivity limits.

\subsubsection{Speckle Imaging}
TOI-1751 was observed on 6 June 2020 using the ‘Alopeke speckle instrument on the Gemini North 8-m telescope \citep{scott+2021}. ‘Alopeke provides simultaneous speckle imaging in two bands (562 nm and 832 nm). Three sets of 1000 $\times$ 0.06~s exposures were collected and subjected to Fourier analysis in the standard reduction pipeline \citep[see][]{howell+2011}. The Fourier transform of the summed autocorrelation of each set of images is used to make a fringe image of the target, which is then used to reconstruct the image. We find no companions fainter than the target star by 4.58 magnitudes at 562 nm and by 6.73 magnitudes at 832 nm at separations of 0.5\arcsec  (i.e. 57 AU) or greater. 

\subsubsection{Adaptive Optics Imaging}
We used the Shane Adaptive optics infraRed Camera-Spectrograph \citep[ShARCS,][]{kupke_et_al2012, gavel_et_al2014} mounted on the 3-m Shane Telescope at Lick Observatory to collect AO imaging of TOI-1751. We observed the target using the $K_s$ and $J$ band filters. We conducted observations using a four-point dither pattern with a spacing of $4\arcsec\,$ on each side. We analyzed the data using the Stellar Image Maturation via Efficient Reduction (SImMER) package \citep{hirsch_et_al2019, savel_et_al2020}. We find no stellar companions within 5 magnitudes at separations  $\geq 1\arcsec\,$ in $K_s$ band.

\begin{figure*}[bth]
    \centering
    \includegraphics[width=0.95\textwidth]{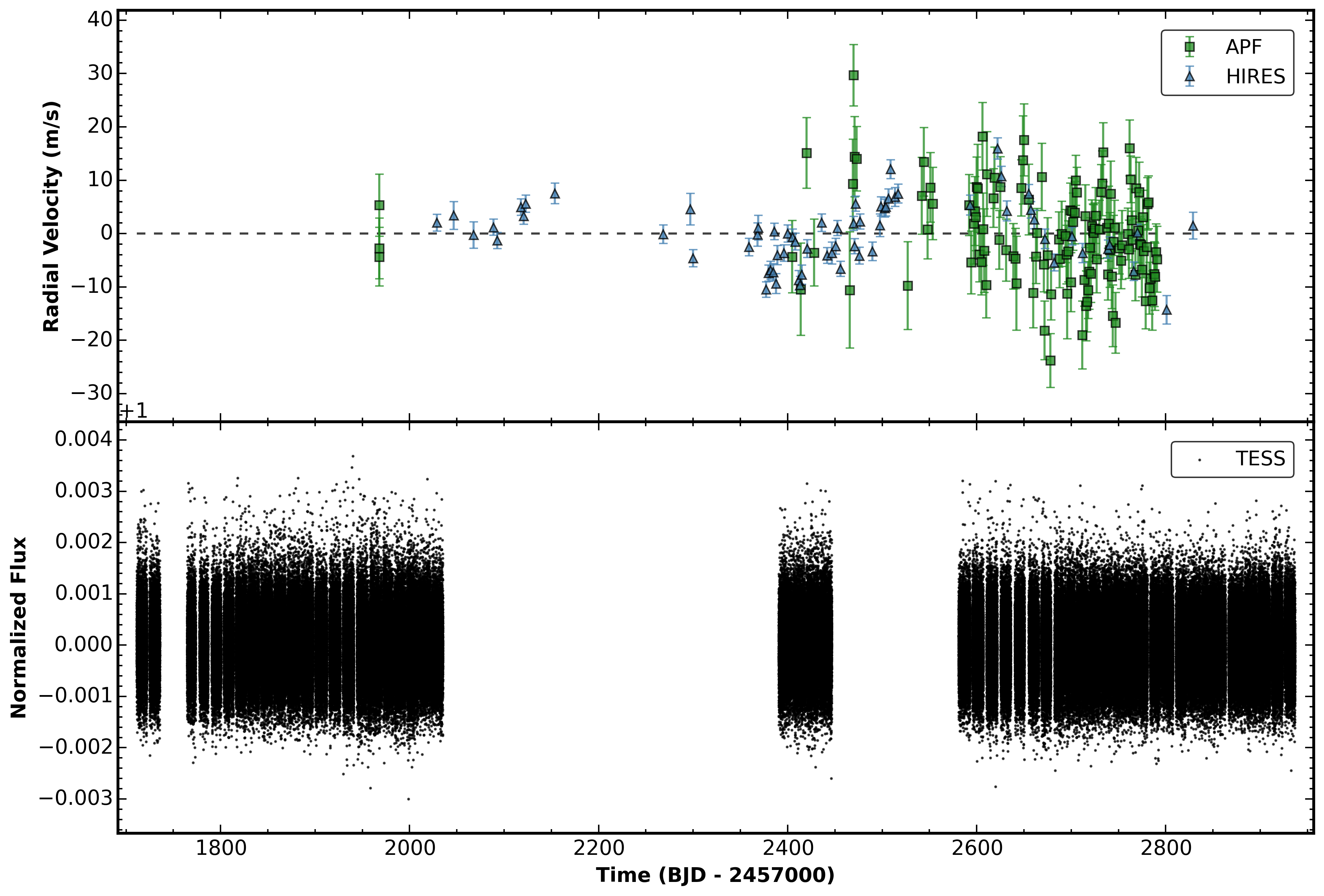}
    \caption{\textbf{Top:} RV measurements from APF/Levy (green squares) and Keck/HIRES (blue diamonds). For clarity, the RV offsets determined from the joint RV and transit fit (see Section \ref{sec:joint}) for each instrument have been subtracted from the data, and the respective jitter terms have been combined with the instrumental uncertainties to produce the error bars. \textbf{Bottom:} Flattened and normalized TESS PDCSAP flux light curve.}
    \label{fig:combo}
\end{figure*}

\subsection{Spectroscopic Observations}

\subsubsection{Reconnaissance Spectra} \label{sec:reconspectra}

We obtained one reconnaissance spectrum with an exposure time of 750~s on 2 March 2020 using the Tillinghast Reflector Echelle Spectrograph (TRES, \citealt{tres}) as part of the TESS Follow-up Observing Program (TFOP) SubGroup 2 (SG2) Reconnaissance Spectroscopy program. TRES is a fiber-fed echelle spectrograph on the 1.5~m Tillinghast Reflector at the Fred Lawrence Whipple Observatory (FLWO) in Arizona, USA, operating between 390 and 910 nm. The spectrograph has a resolving power of R $\sim$ 44,000. We used the Stellar Parameter Classification (SPC, \citealt{buchhave2012}) tool to derive stellar parameters. SPC cross-correlates a $\sim310$ \AA\, region of the observed spectrum surrounding the Mg b lines against a library grid of synthetic spectra calculated using the \cite{kurucz1992} atmospheric models to derive effective temperature ($\rm{T_{eff}}$), surface gravity, ($\rm{log(g)}$), rotational velocity, ($\rm{vsin(i)}$), and metallicity, ([m/H]). Metallicity is derived using all available metal lines and is therefore reported as [m/H]. 

We also obtained follow-up spectra for stellar classification with Las Cumbres Observatory's Network of Robotic Echelle Spectrographs (LCO/NRES, \citealt{Siverd2018}). NRES is a network of four identical spectrographs located at different observatories spanning a wide longitudinal range. Each instrument is a fiber-fed echelle spectrograph operating between 380-860 nm with a resolution of R $\sim$ 53,000. Specifically, we obtained spectra for TOI-1751 at our NRES facility at the Wise Observatory, Israel, on 4 March 2020 at 23:00 UTC. The exposure time was 1800 s and the final SNR was 36 at 5500 \AA. The wavelength calibrated spectrum was obtained through the standard BANZAI-NRES pipeline \citep{banzai-nres}. The resulting stellar parameters from these observations are reported in Table \ref{tab:stellar}.

\subsubsection{Keck/HIRES}
The HIgh Resolution Echelle Spectrometer (HIRES) on the Keck-I telescope on Maunakea operates between 360 and 800 nm. We used Keck/HIRES to collect high-resolution spectra of TOI-1751 in order to derive precision RVs. We collected 71 spectra between July 2020 and July 2023. We observed TOI-1751 using the red cross-disperser, B5 decker (3.5\arcsec $\times$ 0.861\arcsec, R = 50,000) or C2 decker (14\arcsec $\times$ 0.861\arcsec, R = 50,000), and a median exposure time of 382 s. We took 69 RV observations with a warm (50$^{\circ}$ C) iodine cell in the light path for wavelength calibrations as per \citet{butler+1996}. We also took two further higher-resolution spectra without the iodine cell in the light path (``iodine-out'') in March and September 2020 in order to obtain a spectral template, one using the B3 decker (14\arcsec  $\times$ 0.574\arcsec, R = 72,000) and one using the C2 decker. The spectra were reduced using the standard procedures described in \citet{howard+2010}. The RVs, RV errors, and Mount Wilson S-Index (a proxy for stellar activity derived from Ca II H \& K lines) are reported in Table \ref{tab:rvdata}, and the RV data are shown in Fig. \ref{fig:combo}. We also derived stellar parameters for TOI-1751 using these data (see Section \ref{sec:star}, Table \ref{tab:stellar}).

\subsubsection{APF/Levy} \label{sec:apf}
The Automated Planet Finder \citep{radovan+14} is a robotic 2.4-m telescope that hosts the Levy spectrograph, a high-resolution (R=100,000) echelle spectrograph that operates between 374 and 980 nm \citep{ vogt+2014}. We obtained 129 observations of TOI-1751 using APF/Levy between April 2020 and January 2023, with multiple observations per night on several nights. These observations had a median exposure time of 2100~s. Similarly to the Keck/HIRES observations, spectra were taken through a warm iodine cell for wavelength calibration, and RVs were derived using the methods described in \citet{howard+2010}. As with other faint targets on APF (V $\sim 9$), we use the Keck template splined onto the APF wavelength solution to calculate precise RVs. We report the APF/Levy RVs in Table \ref{tab:rvdata}, and show the data in Fig \ref{fig:combo}.

\section{Stellar Parameters} \label{sec:star}
\subsection{Spectroscopically-derived Parameters}
We used the \texttt{SpecMatch-Emp}\footnote{\url{https://github.com/samuelyeewl/specmatch-emp}} algorithm \citep{specmatch-emp} to characterize TOI-1751. \texttt{SpecMatch-Emp} uses a library of high resolution (R$\sim$55,000) and high signal-to-noise ($>$ 100) Keck/HIRES spectra to measure stellar parameters from optical spectra. This method achieves accuracies of 100 K in effective temperature, 15\% in stellar radius, and 0.09 dex in metallicity for FGK stars. The algorithm takes the observed stellar spectrum, shifts it to the rest frame wavelength scale, identifies the most similar library spectra, and interpolates between them to derive parameters for the target star. We measure a stellar mass of $0.90 \pm 0.03 \rm{M_\odot}$, a stellar radius of $1.17 \pm 0.18 \rm{R_\odot}$, and an effective temperature of $5996 \pm 110$ K. We also used the \texttt{SpecMatch-Synthetic}\footnote{\url{ github.com/petigura/specmatch-syn}} code \citep{specmatch-synth} to measure stellar parameters from the iodine-out Keck/HIRES spectrum by interpolating between a grid of model spectra from \citet{coelho+2005}. 

Additionally, we report stellar parameters derived from reconnaissance spectra (described in Section \ref{sec:reconspectra}). However, we do not adopt these parameters in our analysis due to the the higher resolution of Keck/HIRES. Some of these parameters (e.g. $\rm{vsin(i)}$) are discrepant to greater than 2$\sigma$ from those measured using Keck/HIRES data, likely due to the low-resolution and low-cadence (often 1 observation per instrument) nature of these observations. Our adopted parameters are consistent to 1$\sigma$ with those derived homogeneously for the TKS sample in \citet{macdougall+23}.

We find that TOI-1751 is a slightly evolved ($\log \rm{g}  = 4.24 \pm 0.10$), metal-poor ($[\rm Fe/H] = -0.40 \pm 0.06$), solar-type star. The spectroscopic $v\sin i$ and $R_\star$ measured using \texttt{SpecMatch-Emp} imply a projected rotation period for the star of $P_{\rm rot}/\sin i = 46 \pm 37$~d. We derive additional estimates of projected rotation period from the $\rm{log}R^{\prime}_{HK}$ observations, and using the SED of TOI-1751 (see Section \ref{sec:photstar}). We report all calculated stellar parameters in Table \ref{tab:stellar}.

\subsection{Photometrically-derived Parameters} \label{sec:photstar}
We also performed an analysis of the broadband spectral energy distribution (SED) of the star together with the {\it Gaia} DR3 parallax \citep{gaiadr3}, in order to determine an empirical measurement of the stellar radius \citep{Stassun:2016,Stassun:2017,Stassun:2018}. We pulled the $JHK_S$ magnitudes from {\it 2MASS} \citep{2006AJ....131.1163S}, the W1--W4 magnitudes from {\it WISE} \citep{2010AJ....140.1868W}, the $G_{\rm BP} G_{\rm RP}$ magnitudes from {\it Gaia}, and the FUV and NUV magnitudes from {\it GALEX} \citep{2005ApJ...619L...1M}. Together, the available photometry spans the full stellar SED over the wavelength range 0.2--20~$\mu$m.  

We performed a fit using PHOENIX stellar atmosphere models \citep{Husser:2013}, varying effective temperature ($T_{\rm eff}$), metallicity ([Fe/H]), and extinction ($A_V$). We limited $A_V$ to the maximum line-of-sight value from the Galactic dust maps of \citet{Schlegel:1998}. The resulting fit has a best-fit $A_V = 0.03 \pm 0.03$, $T_{\rm eff} = 6075 \pm 75$~K, and [Fe/H] = $-0.5 \pm 0.2$, with a reduced $\chi^2$ of 1.3. Integrating the unreddened model SED gives the bolometric flux at Earth, $F_{\rm bol} = 4.98 \pm 0.23 \times 10^{-9}$ erg~s$^{-1}$~cm$^{-2}$. Taking the $F_{\rm bol}$ together with the {\it Gaia\/} parallax directly gives the bolometric luminosity, $L_{\rm bol} = 2.000 \pm 0.094$~L$_\odot$. The Stefan-Boltzmann relation then gives the stellar radius, $R_\star = 1.284 \pm 0.044$~R$_\odot$. In addition, we estimated the stellar mass and projected stellar rotation period using the empirical relations of \citet{torres+10} ($M_\star = 1.06 \pm 0.06$~M$_\odot$, $\rm{P_{rot}/sin(i)} = 50.9 \pm 40.1~d$). 

We note that the stellar mass derived from the spectroscopic data ($0.90 \pm 0.03 \rm{M_\odot}$) is inconsistent to 1.4 $\sigma$ with that reported in the \tess{} Input Catalog ($1.15 \pm 0.17 \rm{M_\odot} $, TIC v8.2, \citealt{ticv82}), possibly due to the slightly evolved nature of the star. 

We also do not see any evidence of stellar-rotation-related signals in the Lomb-Scargle periodogram \citep{lomb1976, scargle1982} of the \tess{} light curve. The strongest power in the periodogram is at 4.46 d, but given the measured $\rm{v\sin(i)}$ and $\rm{R_*}$ of this target ($1.27 \pm 1.00$ km/s, $1.17 \pm 0.18 \rm{R_*}$), this is unlikely to correspond to stellar rotation, and may be due to \tess{} momentum dumps. We attribute the lack of photometric rotation signals to the known challenges with identifying rotation periods longer than the \tess{} orbital period of 13.7~d, or roughly half a sector length \citep{cantomartins+20, fetherolf+23}. 

\subsection{Constraining Stellar Age and Companions}
To further investigate the rotation period and age of TOI-1751, we use the BANYAN $\Sigma$ (Bayesian Analysis for Nearby Young AssociatioNs $\Sigma$, \citealt{gagne+2018}) analysis tool to investigate whether TOI-1751 is a member of any of the 27 known young stellar associations within 150 pc. We find that TOI-1751 is highly likely ($99.9\%$) to not be a member of these associations, pointing to an older age. This is in line with the reported surface gravity, suggesting that \mbox{TOI-1751} is nearing the end of its time on the main sequence. 

We use the empirical activity-age relations of \citet{mamajek+08} along with the $\log R'_{\rm HK}$ of TOI-1751 to find an age of $10.0 \pm 1.5$~Gyr. This includes both the measurement uncertainty on $\log R'_{\rm HK}$ and the systematic error reported for the \citet{mamajek+08} relation. The same empirical relations of \citet{mamajek+08} predict a rotation period from the $\log R'_{\rm HK}$ of $23.3 \pm 3.2 \rm{d}$. This is consistent with the $P_{\rm rot}/\sin i$ value determined using the \citet{torres+10} empirical relations.

Additionally, Gaia DR3 \citep{gaiadr3} does not contain any proper motion companions within 100\arcsec of TOI-1751. We also used \texttt{tpfplotter} \citep{lillo-box14b} to search for sources in GAIA DR3 within the pipeline aperture mask used to generate the light curve. We found that there were no contaminating sources up to 8 magnitudes fainter than TOI-1751 within the aperture mask, and thus likely no substantial contaminating flux. Furthermore, GAIA DR3 reports a Renormalized Unit Weight Error (RUWE) of 0.90, indicating that this target does not have a detectable companion \citep{ruwe}. 

\subsection{Galactic Context}
TOI-1751 is a high proper motion star, and we find that is also metal-poor ([Fe/H] = $-0.40 \pm 0.06$). These attributes indicate that it may be a member of the thick disk of the Milky Way \citep{bensby04,  bensby10}. Thick disk stars are kinematically hotter, more depleted in metals, more enriched in alpha-elements, and older than those in the thin disk. The formation of the thick disk is still unclear \citep{vandekruit+11}, with possible mechanisms including the merger of the Milky Way with a dwarf galaxy \citep[e.g.]{quinn+93}, and radial mixing of gas and stars \citep[e.g.][]{schonrich+09, loebman+11}. Planets orbiting thick-disk stars (e.g. Kepler-444, \citealt{campante+15}) show that planet formation has occurred for $\gtrsim11$ Gyr. \citet{carrillo+20} calculated the galactic velocity of TIC stars, and report $\rm{(U_{LSR}, V_{LSR}, W_{LSR})}$ = $\rm{(106.2 \pm 0.2, -35.5 \pm 0.2, 1.6 \pm 0.2)}$ km/s, with $\rm{v_{tot} = (U_{LSR}^2 + V_{LSR}^2 + W_{LSR}^2)^\frac{1}{2} \approx 110 km/s}$ for this target. This result is consistent with that calculated using the methods in \citet{rodriguez+16}, and may suggest thick-disk membership \citep{nissen+04}. TOI-1751 is, however, still consistent with the low-metallicity tail of the thin-disk distribution, and is 1.6 times more likely to belong to the thin disk than the thick disk \citep{carrillo+20}. Therefore, we do not conclusively report its thick disk membership, and defer a detailed discussion of the star's alpha abundances to a future work (Polanski et al., in prep).

\startlongtable 
\begin{deluxetable*}{ccccc}
\tablecaption{TOI-1751 Stellar Parameters \label{tab:stellar}}
\tablehead{\colhead{Parameter} & \colhead{Value} & \colhead{Error} & \colhead{Source} & \colhead{Adopted?}} 
\startdata
Other Names & TIC 287080092 & & TIC v8.2\tablenotemark{a} & - \\[-0.05cm]
& HD 146757 & & Henry Draper Catalog \citep{cannon+1921} & - \\[-0.05cm]
& TYC 4192-02025-1 & & TYCHO \citep{tycho} & - \\[-0.05cm]
Right Ascension (hh:mm:ss) & 16:13:57.31 & &  TIC v8.2\tablenotemark{a} & - \\[-0.05cm]
Declination (hh:mm:ss) & +63:32:03.39 & & TIC v8.2\tablenotemark{a} & - \\[-0.05cm]
V magnitude & 9.327 & 0.003 & TIC v8.2\tablenotemark{a} & - \\[-0.05cm]
TESS magnitude & 8.80616 & 0.006 & TIC v8.2\tablenotemark{a} & - \\[-0.05cm]
J magnitude & 8.251 & 0.021 & TIC v8.2\tablenotemark{a} & - \\[-0.05cm]
K magnitude & 7.934 & 0.027 & TIC v8.2\tablenotemark{a} & - \\[-0.05cm]
\textit{Gaia} magnitude & 9.19 & - & \textit{Gaia} DR3\tablenotemark{b} & - \\[-0.05cm]
Parallax (mas) & 8.809 & 0.009 & \textit{Gaia} DR3\tablenotemark{b} & - \\[-0.05cm]
RA proper motion (mas/yr) & 8.60 & 0.01
	 & \textit{Gaia} DR3\tablenotemark{b}  & - \\[-0.05cm]
Dec proper motion (mas/yr) & -172.84 & 0.02 & \textit{Gaia} DR3\tablenotemark{b}  & - \\
\hline
Radius $\rm{(R_\odot)}$ & 1.17  & 0.18 &  SpecMatch-Empirical\tablenotemark{c} &  Y \\[-0.05cm]
Radius $\rm{(R_\odot)}$ & 1.34  & 0.03 &  SpecMatch-Synthetic\tablenotemark{d} &   \\[-0.05cm]
Radius $\rm{(R_\odot)}$ & 1.27
& 0.06 &  TIC v8.2\tablenotemark{a} & - \\[-0.05cm]
Radius $\rm{(R_\odot)}$ & 1.01
& 0.11 &  LCO/NRES & - \\[-0.05cm]
Radius $\rm{(R_\odot)}$ & 1.284
& 0.044 & SED & - \\
\hline
Mass $\rm{(M_\odot)}$ & 0.90 & 0.03  &  SpecMatch-Empirical\tablenotemark{c} & Y \\[-0.05cm]
Mass $\rm{(M_\odot)}$ & 0.89 & 0.03  &  SpecMatch-Synthetic\tablenotemark{d} & \\[-0.05cm]
Mass $\rm{(M_\odot)}$ & 1.152
& 0.1689 &  TIC v8.2\tablenotemark{a} & - \\[-0.05cm]
Mass $\rm{(M_\odot)}$ & 0.925
& 0.044 &  LCO/NRES & - \\[-0.05cm]
Mass $\rm{(M_\odot)}$ & 1.06
& 0.06 & Empirical relations \citep{torres+10} & - \\
\hline
$\rm{T_{eff} (K)}$ & 5996 & 110 &  SpecMatch-Empirical\tablenotemark{c} & Y \\[-0.05cm]
$\rm{T_{eff} (K)}$ & 5918 & 100 &  SpecMatch-Synthetic\tablenotemark{d} & - \\[-0.05cm]
$\rm{T_{eff} (K)}$ & 6114 & 122 &  TIC v8.2\tablenotemark{a} & - \\[-0.05cm]
$\rm{T_{eff} (K)}$ & 5850 & 50 &  NOT/FIES  & - \\[-0.05cm]
$\rm{T_{eff} (K)}$ & 5801 & 50 &  FLWO/TRES  & - \\[-0.05cm]
$\rm{T_{eff} (K)}$ & 6049 & 100 &  LCO/NRES  & - \\[-0.05cm]
$\rm{T_{eff} (K)}$ & 6075 & 75 &  SED  & - \\
\hline
$\rm{log(g)}$ & 4.24 & 0.10 &  SpecMatch-Synthetic\tablenotemark{c}  & Y \\[-0.05cm]
$\rm{log(g)}$ & 4.293 & 0.084 & TIC v8.2\tablenotemark{a} & - \\[-0.05cm]
$\rm{log(g)}$ & 4.1 & 0.1 &  NOT/FIRES & - \\[-0.05cm]
$\rm{log(g)}$ & 4.0 & 0.1 &  FLWO/TRES  & - \\[-0.05cm]
$\rm{log(g)}$ & 4.4 & 0.1 &  LCO/NRES  & - \\
\hline
$\rm{v sin\textit{i}}$ (km/s) & 1.27 & 1.0  & SpecMatch-Empirical\tablenotemark{c} &  Y \\[-0.05cm]
$\rm{v sin\textit{i}}$ (km/s) & 3.74 & 0.50 &  NOT/FIRES & - \\[-0.05cm]
$\rm{v sin\textit{i}}$ (km/s) & 4.29 & 0.5 &  FLWO/TRES  & - \\[-0.05cm]
$\rm{v sin\textit{i}}$ (km/s) & 2.49 & 0.62 &  LCO/NRES  & - \\
\hline
$\rm{P_{rot}/sin(i) (d)}$ & 46 & 37 & Calculated using $\rm{vsin\textit{i}}$ and $\rm{R_*}$ & - \\[-0.05cm]
$\rm{P_{rot}/sin(i) (d)}$ & 50.9 & 40.1 & Empirical relations \citep{torres+10} & - \\[-0.05cm]
$\rm{P_{rot} (d)}$ & 23.3 & 3.2 & Predicted using  $\rm{log}R^{\prime}_{HK}$ \citep{mamajek+08}& - \\
\hline
$\rm{{[Fe/H]}~(dex)}$ & -0.40 & 0.06  & SpecMatch-Empirical\tablenotemark{c} & Y \\[-0.05cm]
$\rm{{[Fe/H]}~(dex)}$ & -0.50 & 0.09  & SpecMatch-Synthetic\tablenotemark{c} & - \\[-0.05cm]
$\rm{{[Fe/H]}~(dex)}$ & -0.33 & 0.06 &  LCO/NRES  & - \\[-0.05cm]
$\rm{{[Fe/H]}~(dex)}$ & -0.5 & 0.2 &  SED  & - \\
\hline
$\rm{log}R^{\prime}_{HK}$ & -5.244 & 0.212 & APF/Levy & - \\[-0.05cm]
$\rm{log}R^{\prime}_{HK}$ & -5.149 & 0.071 & Keck/HIRES & - \\ 
$\rm{{[m/H]}~(dex)}$ & -0.429 &  0.08 &  NOT/FIRES & - \\[-0.05cm]
$\rm{{[m/H]}~(dex)}$ & -0.389 & 0.08 &  FLWO/TRES  & - \\[-0.05cm]
\hline 
$\rm{A_V}$ & 0.03 & 0.03 & SED & - \\[-0.05cm]
Age (Gyr) & 10.0 & 1.5 & Empirical relations \citep{mamajek+08} & - 
\enddata
\tablenotetext{a}{ \tess{} Input Catalog, Version 8.2 \citep{ticv82}}
\tablenotetext{b}{\textit{Gaia} DR3 \citep{gaiadr3}} 
\tablenotetext{c}{SpecMatch-Empirical \citep{specmatch-emp} applied to Keck/HIRES data} 
\tablenotetext{d}{SpecMatch-Synthetic \citep{specmatch-synth} applied to Keck/HIRES data}
\end{deluxetable*}

\section{Data Analysis} \label{sec:analysis}
\subsection{TESS Photometry Analysis}\label{sec:tess_photometry}

We used the \texttt{lightkurve} package \citep{Lightkurve_2018} to download the 2-minute SPOC TESS light curves for TOI-1751.  We normalized and stitched together the light curves from each TESS Sector (see Fig. \ref{fig:combo}) and computed a box least-squares (BLS, \citealt{kovacs+2002A&A}) periodogram. We recovered a significant periodic signal at 37.468 d with an associated Signal Detection Efficiency (SDE) of 24.8. This signal corresponds to the planet candidate \mbox{TOI-1751.01.}

We visually inspected the light curve to confirm the time of the first transit of TOI-1751.01. Next, we reduced the computation time for the subsequent transit fit using the BLS period to trim data points falling outside of a two-day window on either side of each transit center. We also removed any observations flagged with a quality flag greater than 0 to exclude scattered light, cosmic rays, and additional anomalous events.

Next, we flattened each transit individually by first fitting a second-order polynomial to the out-of-transit baseline flux spanning 12 to 48 hours before and after each transit midpoint. We then divided out the best-fit polynomial from the Presearch Data Conditioning Simple Aperture Photometry (PDCSAP, \citealt{2012PASP..124.1000S, Stumpe2012, Stumpe2014}) flux within the full four-day window for each transit. To ensure robust flattening, we required each four-day window to be at least 80\% complete (i.e., at least 80\% of possible cadences contained data). This resulted in 13 complete flattened transits.

To constrain the size and orbital properties of TOI-1751~b, we modeled the transit photometry using the \texttt{exoplanet} package \citep{exoplanet:joss}. To remain agnostic to the stellar properties, we defined a transit model in terms of the planet-to-star radius ratio ($\rm{R_p/R_*}$), time of first transit ($\rm{T_0}$), orbital period ($\rm{P}$), semi-major axis in units of stellar radii ($\rm{a/R_*}$), impact parameter (b), and quadratic limb-darkening coefficients \citep[$\rm{q_1}$, $\rm{q_2}$, using the parameterization of][]{kipping+2013MNRAS}. Our model also included a mean baseline flux term ($\langle F \rangle$) and a photometric jitter term (s), which was added in quadrature to the reported flux uncertainties. Finally, we assumed a circular orbit for the planet, setting the eccentricity equal to zero.

We optimized the model parameters using Bayesian inference, implementing a Hamiltonian Monte Carlo (HMC) No U-Turn Sampler \citep[NUTS,][]{Hoffman+Gelman_2011} with \texttt{PyMC3} \citep{exoplanet:pymc3} to sample the posterior probability distributions. The prior distributions we selected for each parameter are given in Table \ref{tab:distributions}.  We set the target acceptance rate to 0.95 (to account for the higher acceptance fractions returned by HMC samplers compared with Metropolis-Hastings samplers) and initialized the sampler by adapting a dense mass matrix from the sample covariances. We then ran the sampler using a total of 4 chains, each one drawing 20,000 samples after discarding 5,000 burn-in steps. To check for convergence, we computed the Gelman-Rubin Diagnostic and visually inspected the sampler trace plot for each parameter. The median and 68\% confidence range for each parameter are given in Table \ref{tab:distributions}.

\begin{figure}[bth]
    \centering
    \includegraphics[width=0.9\columnwidth]{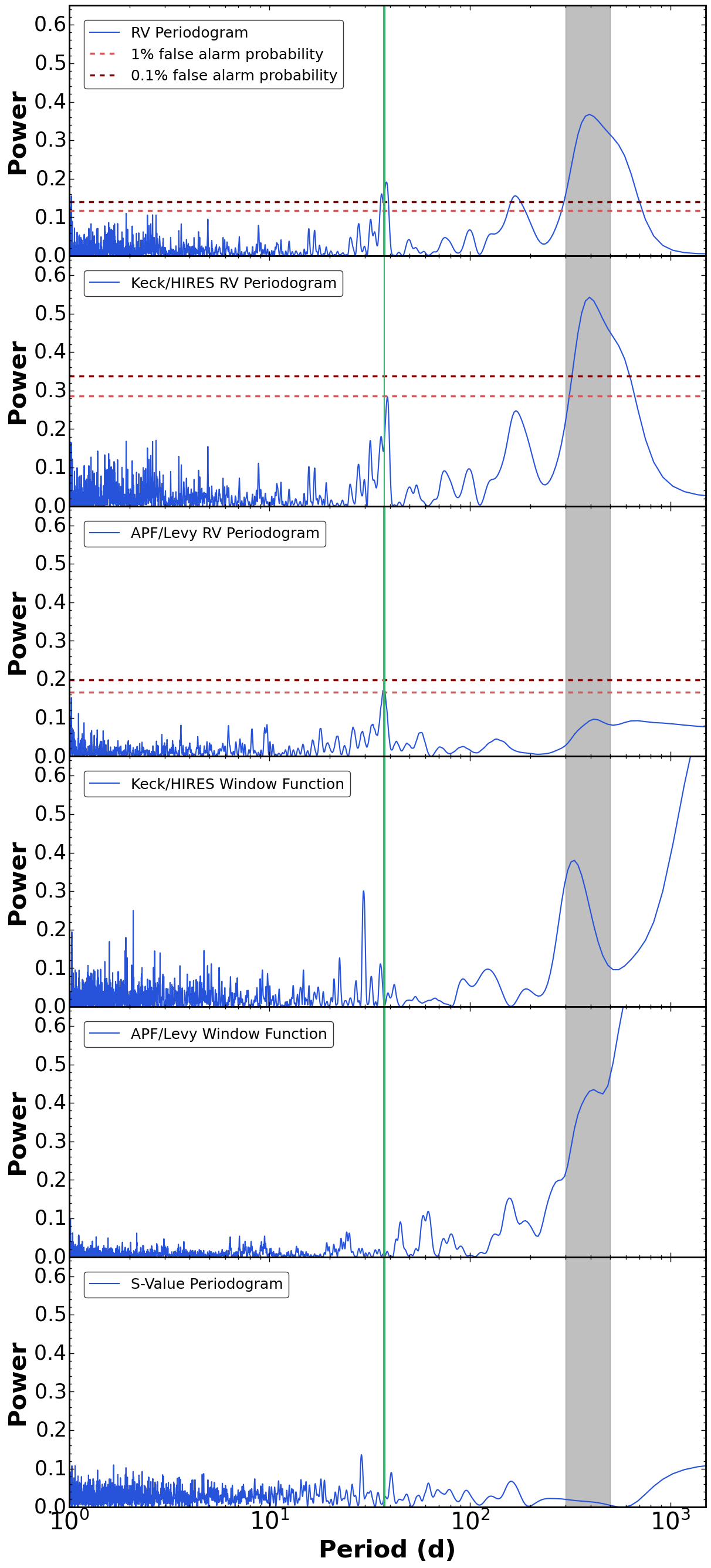}
    \caption{Lomb-Scargle periodogram of a) all RV observations, b) Keck/HIRES RVs, c) APF/Levy RVs, d) Keck/HIRES window function, e) APF/Levy window function, and f) S-values. The vertical green line marks the 37d period, and the gray-shaded region marks the 300-500d range. The long-period peak seen in the periodogram of the full RV data set is not present in the APF data, and is at a period similar to that of a significant peak in the HIRES window function. The dark red dashed line represents the 0.1\% false-alarm level. Data consisting of Gaussian noise with no periodic signal would produce peaks of this height (or above) in $<0.1\%$ of samples. The red dashed line is the 1\% false-alarm level.}
    \label{fig:periodogram}
\end{figure}

\subsection{RV Analysis} \label{sec:rv_analysis}
We analyze the combined Keck/HIRES and APF/Levy RV observations in order to confirm TOI-1751.01. The Lomb-Scargle periodogram of the combined Keck/HIRES and APF/Levy RV data (see Fig. \ref{fig:periodogram}) shows a significant peak with $<0.1\%$ false alarm probability at 37.4 d, which we confirm as the exoplanet TOI-1751~b. We also note a peak at $\sim$ 400 d. This second, longer-period signal may indicate the presence of a non-transiting distant giant planet in the system. 

We used the \texttt{RadVel} package \citep{radvel} which fits Keplerian models using maximum posterior probability optimization to model the RV data. In our models, we allowed several combinations of the orbital period (P), time of conjunction ($t_c$), mass ($M_{pl}$), argument of periastron ($\omega$), and eccentricity (e) of the planet(s) to vary. We performed several fits: including or excluding a linear trend ($\dot{\gamma}$), including or excluding a second planet at $\sim400$ d, and circular or eccentric orbits. In all cases, we fixed the orbital period ($P_b$) and time of conjunction ($t_{c,b}$) using the precise constraints from our initial photometric analysis (see Section \ref{sec:tess_photometry}). 

We find that the mass of TOI-1751~b is consistent to $1\sigma$ across all models. Furthermore, the mass is consistent to $1\sigma$ when fitting the Keck/HIRES and APF/Levy data separately. The preferred model using the full data set is a circular 1-planet model, which returns a minimum mass of \textbf{\Mpl{}} for TOI-1751~b. The $\rm{\Delta BIC}$ between the circular model and eccentric model is less than 1, so there is no clear evidence for eccentricity \citep{kass1995bayes}. We report the circular posteriors in Table \ref{tab:distributions}.

\subsubsection{Evidence Against a Distant Giant Planet}
There is much interest in probing the outer regions of planetary systems, and investigating the occurrence of small, inner planets that may depend on the presence of distant giants \citep[e.g.][]{vanzandt+23, zhu+wu18}. We find that the $\sim$ 400 d periodic signal seen in the combined data set is only present in the Keck/HIRES RVs, and is not seen in the APF/Levy data (see Fig. \ref{fig:periodogram}). This casts doubt on a planetary interpretation of the 400 d signal. 

\begin{figure}[bth]
    \centering
    \includegraphics[width=\columnwidth]{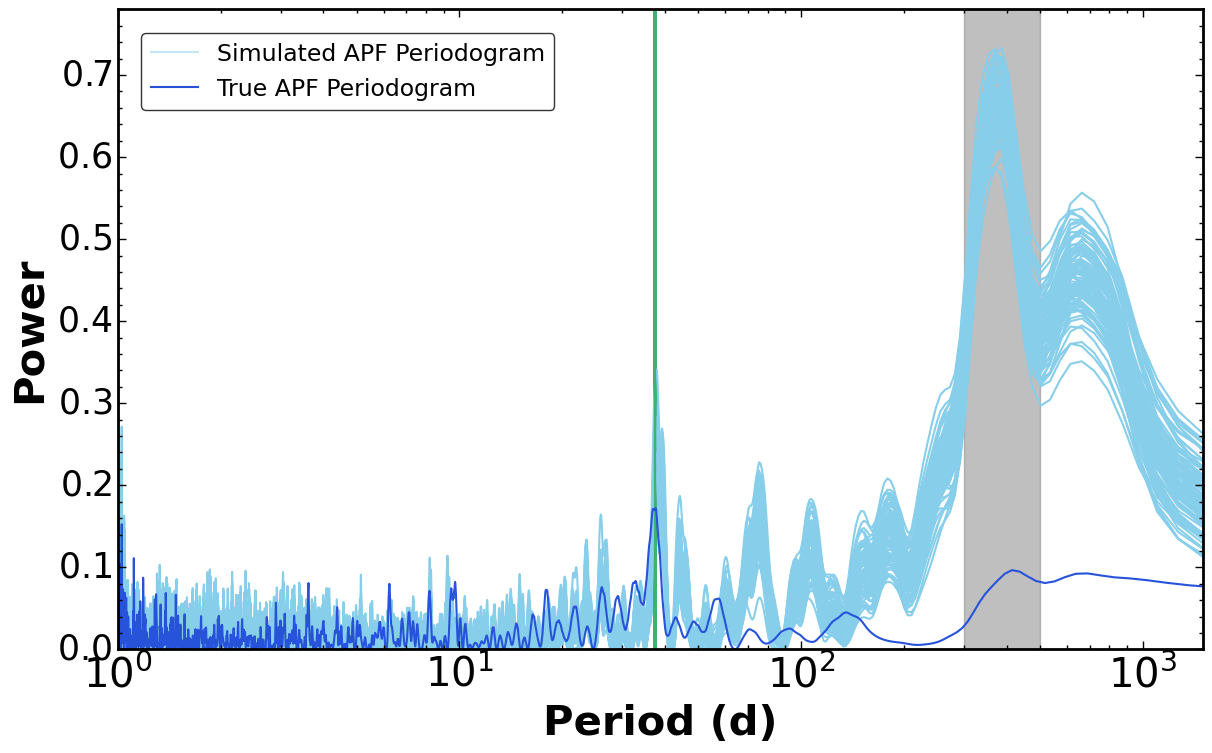}
    \caption{Lomb-Scargle periodogram of APF/Levy RV data (blue), and 100 simulations of APF/Levy observations of a 2-planet TOI-1751 system. The period of TOI-1751~b is indicated by a vertical green line. The peaks in the periodogram of the simulated data are consistently higher than those of the real data at $\sim 400$ d.}
    \label{fig:apf_sims}
\end{figure}

In order to investigate whether a distant giant planet would be detected in the APF/Levy data, we simulated observations using parameters for TOI-1751~b and the putative outer planet determined from a two-planet Keck/HIRES-only fit. We used M = $14.9 \pm 3.8 M_\oplus $ for the inner planet and M = $88 \pm 15 M_\oplus$ ($\sim$0.9 $M_{Saturn}$) for the outer planet to simulate 129 observations, using real APF/Levy observational time stamps and errors. All of the simulated observations show a long-period peak greater than that in the real APF/Levy data (see Fig. \ref{fig:apf_sims}). This suggests that if such a distant giant planet were present in the system, its signal would have been captured by the APF observations. We performed a similar exercise for simulated APF data assuming lower masses (10, 35, and 50 $\rm{M_\oplus}$) for a putative distant giant. None of these scenarios reproduced the periodogram of the APF data. We used \texttt{Radvel} to model these sets of simulated observations. The measured planet mass for the outer planet is inconsistent with the injected values. As such, we did not find a planetary solution for the 400 d signal that is consistent with observations of TOI-1751~b, and conclude that this long-period signal is not planetary in nature. 

A possible cause of this spurious signal is that the stellar template used in reducing the RV observations was sub-optimal. This would cause a correlation between the barycentric velocity of the observations and the RV measurements. However, we would expect such an effect to impact the APF/Levy observations as well. We also computed the spectral window function of our observations, i.e. the Lomb-Scargle periodogram of observation times. There is substantial power at $\sim200$ d in the window function of the Keck/HIRES observations, indicating that the periodic signal in the RV data may be a harmonic of a signal in the window function due to patterns in observation times. Given that the $\sim 400$ d periodic signal is only present in the Keck/HIRES data, that it would be detectable in the APF/Levy data assuming it was astrophysical, and that there is a corresponding peak in the Keck/HIRES window function, we conclude that this signal is not planetary in nature. This highlights the valuable role that the APF, and other comparable telescopes, can play in large RV surveys. High-cadence APF/Levy observations of TOI-1751, despite being lower precision than the Keck/HIRES data, allowed us to identify and remove an alias in our data, and to gain a clearer picture of this system.

\subsection{Joint Photometry and RV Analysis} \label{sec:joint}
Using the results from the previous two sections as a baseline, we performed a joint fit of the 13 flattened TESS transits described in Section \ref{sec:tess_photometry} and the APF and HIRES RV measurements described in Section \ref{sec:rv_analysis}. 

As in Section \ref{sec:tess_photometry}, we used the \texttt{exoplanet} package \citep{exoplanet:joss}. We define the combined photometry and RV model by a set of 16 free parameters: $R_p/R_*$, $T_0$, $P$,  $b$, $q_1$, $q_2$, and $\langle F \rangle$ (as in the transit-only model), as well as stellar mass and radius, $M_*$ and $R_*$, and planet mass $M_p$. We also included three jitter terms (one per instrument), which were added to the respective instrument uncertainties in quadrature, as well as RV offsets, $\gamma$, for both APF and HIRES. Finally, we added an RV acceleration term, $\dot{\gamma}$. We also derived the following quantities from the fit parameters: the stellar density ($\rm{\rho_*}$), planet radius ($\rm{R_p}$), planet bulk density ($\rm{\rho_p}$), RV semi-amplitude (K), transit duration ($\rm{t_{dur}}$), and semi-major axis (a). As stated in Section \ref{sec:rv_analysis}, the $\rm{\Delta BIC}$ between circular and eccentric is less than 1, indicating no clear evidence for eccentricity, so we adopt the circular model posteriors. We report both circular and eccentric model posteriors in Rable \ref{tab:distributions} for completeness.

\begin{figure}[bth]
    \includegraphics[width=\columnwidth]{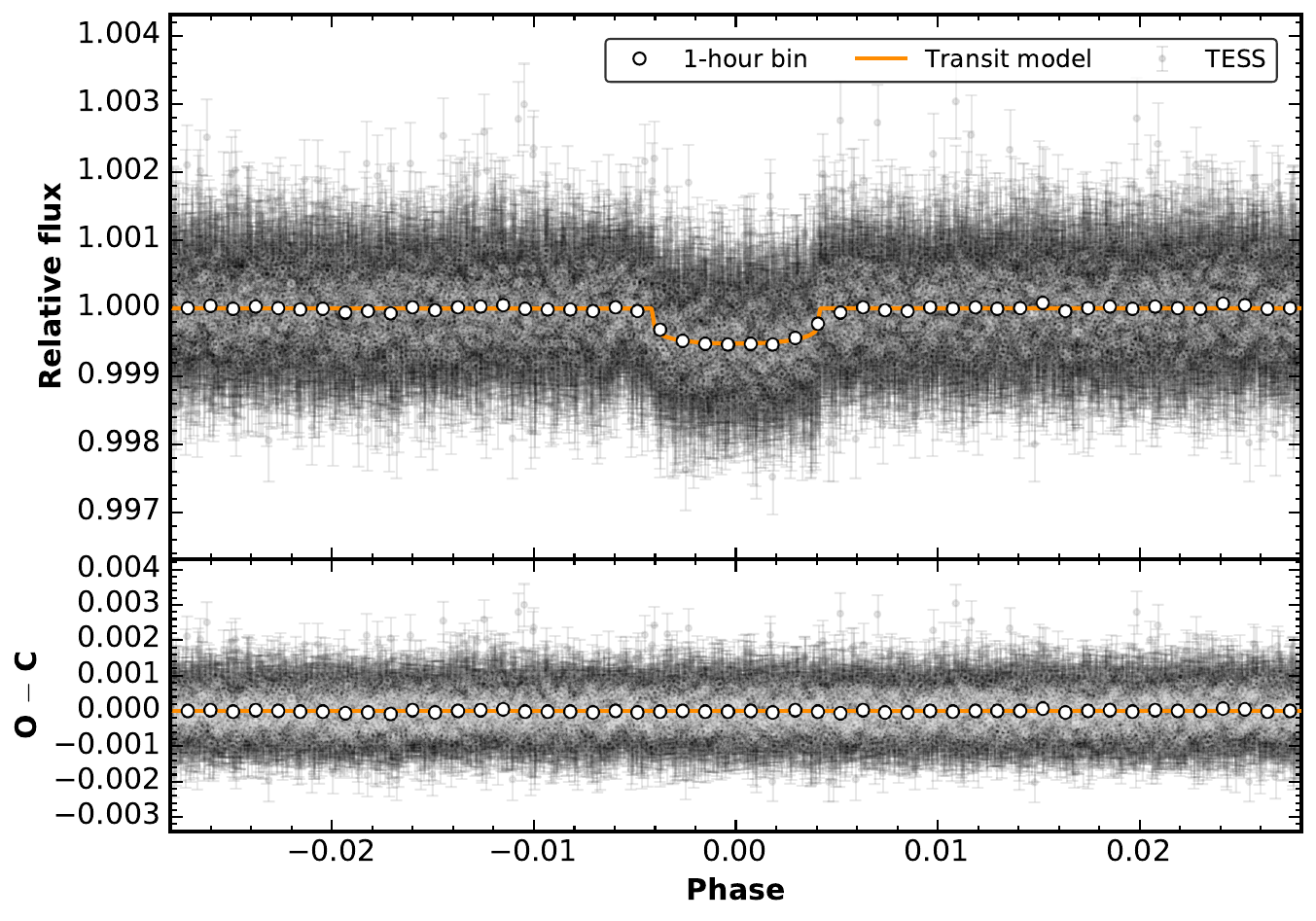}
    \caption{\textbf{Top:} Phase-folded TESS light curve of TOI-1751 b. The black points with error bars show the flattened PDCSAP flux from 13 transits, where the error bars are calculated by adding the jitter and instrument uncertainty in quadrature. The orange line shows the best-fit (median) transit model from the joint RV and transit analysis. For clarity, the data are also shown in 1-hour bins (white circles). \textbf{Bottom:} Residual flux between the best-fit (median) transit model and the data.}
    \label{fig:joint_transit_folded}
\end{figure}

\begin{figure*}[bth]
    \centering
    \includegraphics[width=0.95\textwidth]{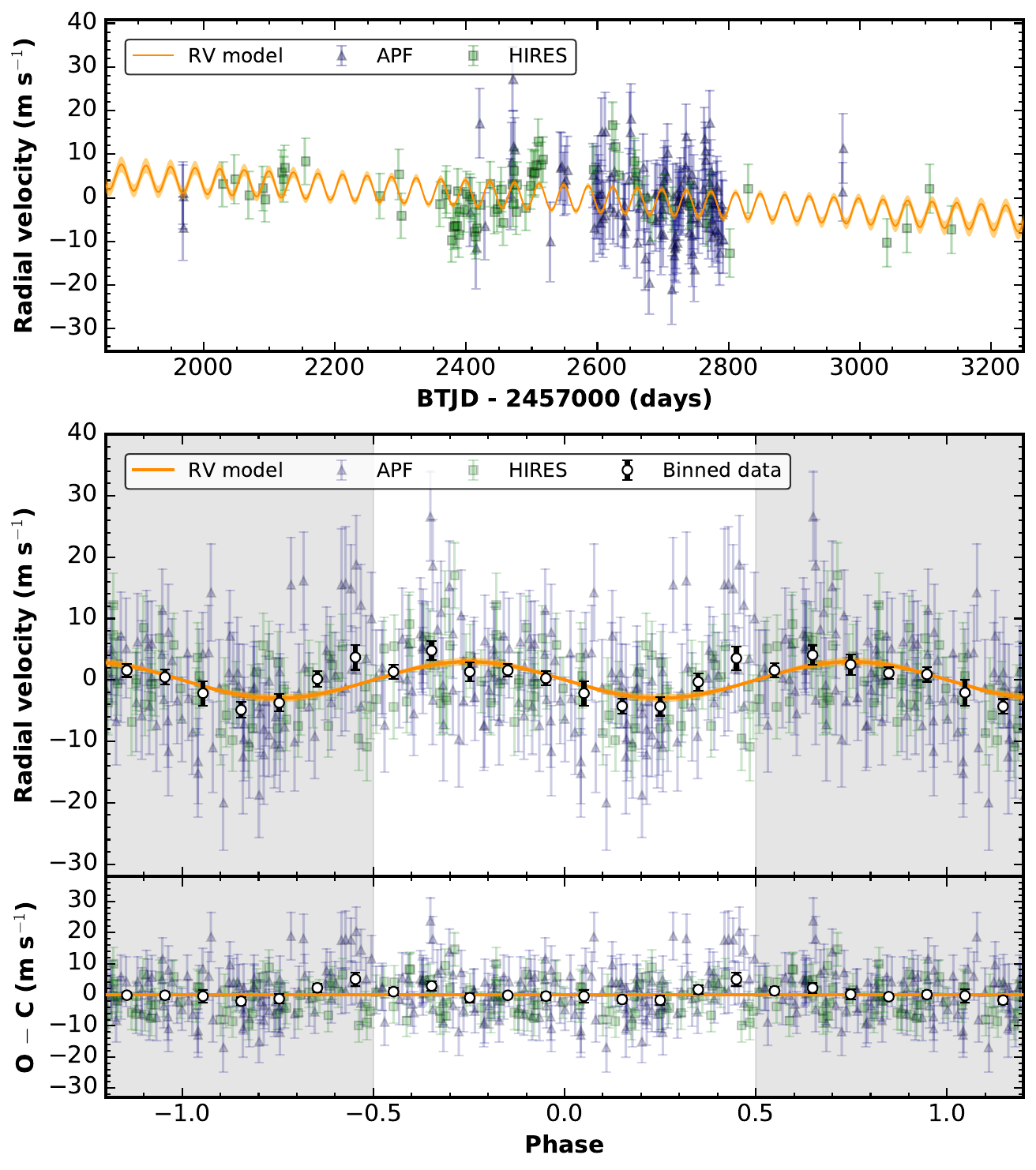}
    \caption{\textbf{Top:} RV measurements from APF (blue diamonds) and HIRES (green squares) and the best-fit (median) RV model (orange line) from the joint RV and transit analysis. For clarity, the RV offset for each instruments has been subtracted from the data, and the respective jitter terms have been combined with the instrumental uncertainties to produce the error bars. The orange shaded region shows the 68\% confidence range for the model. \textbf{Middle:} Phase-folded radial velocities of TOI-1751 b, with the instrumental offsets and linear trend removed. The APF data (blue triangles) and HIRES data (green squares) are shown with error bars and the best-fit (median) RV model for the planet (orange line). The jitter term for each instrument has been added to the RV uncertainty in quadrature to calculate the error bars. The orange shaded region indicates the model 68\% confidence range. For clarity, repeated orbital phase coverage is shown in the gray shaded regions, and we have also binned the data to $\sim$4 d intervals (white circles with error bars). \textbf{Bottom:} Residual radial velocities between the best-fit (median) model and the data.}
    \label{fig:joint_RVs_folded}
\end{figure*}

Again, we used the HMC NUTS implemented in \texttt{PyMC3} \citep{exoplanet:pymc3} to optimize the model parameters and sample their posterior probability distributions. The priors we adopted for each of these parameters are shown in Table \ref{tab:distributions}. We followed the procedure described in Section \ref{sec:tess_photometry} to initialize the sampler and sample the posteriors. The full posterior distributions for the joint transit and RV fit are shown in Figure \ref{fig:joint_b_posteriors} in the Appendix, and the posteriors for the auxiliary parameters are shown in Figure \ref{fig:joint_b_posteriors_aux}. Finally, phase-folded plots of the data and models are shown in Figures \ref{fig:joint_transit_folded} and \ref{fig:joint_RVs_folded}. The median posterior values for each parameter, along with 68\% confidence intervals, are shown in Table \ref{tab:distributions}. We confirm the planetary nature of TOI-1751~b, and measure its radius (\Rpl{}), mass (\Mpl{}), and bulk density (\rhopl{}). 

\section{Discussion} \label{sec:discussion}
\subsection{TOI-1751~b In Context}
Looking to our Solar system, the ice giant planets Uranus and Neptune are the most similar to TOI-1751~b in mass. However, their internal compositions, gravitational fields, rotation periods, and atmospheric dynamics are poorly constrained \citep{podolak+12, neuenschwander+22, miguel+23}. These substantial uncertainties about ice giant interiors inhibit the use of Solar System benchmarks to inform models of extra-solar planets. 

\vspace{-10cm}

\begin{longrotatetable}
\movetabledown=10mm
\begin{deluxetable*}{LCCCCCCCC}
\tablecolumns{9}
\tablefontsize{\scriptsize}
\tablecaption{Prior distributions and posterior quantiles derived from model fits.}
\label{tab:distributions}
\tablehead{
    \colhead{} &
    \multicolumn{2}{C}{\textbf{RV-only fit}} &
    \multicolumn{2}{C}{\textbf{Transit-only fit}} &
    \multicolumn{2}{C}{\textbf{RV+transit fit (eccentric)}}  &
    \multicolumn{2}{C}{\textbf{RV+transit fit (circular)}} \\
    \colhead{Parameter} &
    \colhead{Prior} &
    \colhead{$^{*}$Posterior median} &
    \colhead{Prior} &
    \colhead{$^{*}$Posterior median} &
    \colhead{Prior} &
    \colhead{$^{*}$Posterior median} &
    \colhead{Prior} &
    \colhead{$^{*}$Posterior median}
}
\startdata
\textbf{\textit{Stellar parameters}}                            & & & & & & & & \\
Stellar mass, $M_*$ ($M_{\odot}$)                                 & - & 0.90 $\pm$  0.03 & - & - & $\mathcal{T}(0.90, 0.03, 0, 3)$ & 0.90 $\pm$  0.03 & $\mathcal{T}(0.90, 0.03, 0, 3)$ & 0.90 $\pm$ 0.03\\
Stellar radius, $R_*$ ($R_{\odot}$)                               & - & - & - & - & $\mathcal{T}(1.17, 0.18, 0, 3)$ & $1.21_{-0.12}^{+0.12}$ & $\mathcal{T}(1.17, 0.18, 0, 3)$ & $1.20_{-0.03}^{+0.06}$\\
Stellar bulk density, $\rho_{*}$ ($g cm^{-3}$)        & - & - & - & - & - & $0.72_{-0.18}^{+0.26}$& - & $0.74_{-0.10}^{+0.05}$ \\
Mean flux, $\langle F \rangle$                                  & - & - & $\mathcal{N}(1.00, 0.01)$ & $0.999998 \pm 0.000003$ & $\mathcal{N}(1.00, 0.01)$ & $0.999998 \pm 0.000003$ & $\mathcal{N}(1.00, 0.01)$ & $0.999998 \pm 0.000003$ \\
$^{**}$Quadratic limb darkening coeff., $q_{1}$                   & - & - & $\mathcal{U}(0, 1)$ & $0.30_{-0.17}^{+0.14}$ & $\mathcal{U}(0, 1)$ & $0.29_{-0.16}^{+0.14}$ & $\mathcal{U}(0, 1)$ & $0.29_{-0.16}^{+0.14}$ \\
$^{**}$Quadratic limb darkening coeff., $q_{2}$                   & - & - & $\mathcal{U}(0, 1)$ & $0.26_{-0.19}^{+0.28}$ & $\mathcal{U}(0, 1)$ & $0.24_{-0.17}^{+0.26}$ & $\mathcal{U}(0, 1)$ & $0.24_{-0.17}^{+0.26}$ \\
\textbf{\textit{Planet parameters}}                             & & & & & & & & \\
Planet-to-star radius ratio, $R_{p} / R_{*}$                    & - & - & $\mathcal{L}(-3.86, 0.01)$ & 0.02110 $\pm$  0.00019 & $\mathcal{N}(0.0211, 0.0002)$ & 0.02119 $\pm$  0.00017 & $\mathcal{N}(0.0211, 0.0002)$ & 0.02119 $\pm$ 0.00017 \\
Planet radius, $R_{p}$ ($R_{\oplus}$)                                  & - & - & - & - & - & $2.79_{-0.27}^{+0.29}$ & - & $2.77_{-0.07}^{+0.15}$ \\
Planet mass, $M_{p}$ ($M_{\oplus}$)                                    & $>$ 0 & $17.6 \pm 3.5$ & - & - & $\mathcal{L}(2.8, 2.0) $ & $17.1 \pm 3.2$ & $\mathcal{L}(2.8, 2.0)$ & $14.5_{-3.14}^{+3.15}$ \\
Planet bulk density, $\rho_{p}$ ($g cm^{-3}$)             & - & - & - & - & - & $4.2_{-1.3}^{+1.8}$ & - & $3.6 \pm 0.9$ \\
\textbf{\textit{Orbital parameters}}                            & & & & & & & & \\
Orbital period, $P$ (days)                                      & $=37.468$ & 37.468 & $\mathcal{L}(3.62, 0.01)$ & $37.46850_{-0.00009}^{+0.00011}$ & $\mathcal{L}(3.62, 0.01)$ & $37.468489_{-0.000074}^{+0.000082}$ & $\mathcal{L}(3.62, 0.01)$ & $37.468490_{-0.000075}^{+0.000082}$ \\
First transit center, $T_{0}$ ($BJD - 2457000$)           & - & - & $\mathcal{N}(1733.6, 0.1)$ & $1733.6349_{-0.0025}^{+0.0019}$ & $\mathcal{N}(1733.635, 0.0001)$ & $1733.6351_{-0.0015}^{+0.0013}$ & $\mathcal{N}(1733.635, 0.0001)$ & $1733.6352_{-0.0015}^{+0.0014}$ \\
Transit duration, $T_{dur}$ (hr)                           & - & - & - & - & - & $9.5_{-1.3}^{+1.4}$ & - & $7.5_{-0.06}^{+0.08}$ \\
Semi-major axis, $a/R_{*}$                                    & - & - & $\mathcal{U}(10, 60)$ & $37.92_{-1.97}^{+0.86}$ & - & $37.6_{-3.5}^{+4.0}$ & - & $38.0_{-1.8}^{+0.8}$ \\
Semi-major axis, $a$ (AU)                                       & - & $0.2116^{+0.0023}_{-0.0024}$ & - & - & - & $0.2115_{-0.0024}^{+0.0023}$ & - & $0.2116 \pm 0.0023$ \\
Impact parameter, $b$                                           & - & - & $\mathcal{U}(0, 1)$ & $0.20_{-0.13}^{+0.17}$ & $\mathcal{U}(0, 1)$ & $0.20_{-0.14}^{+0.16}$ & $\mathcal{U}(0, 1)$ & $0.20_{-0.14}^{+0.16}$ \\
Eccentricity, $e$                                               & $=0.0$ & 0.0 & 0.0 & 0.0 & $\mathcal{U}(0, 1)$ & $0.261_{-0.090}^{+0.077}$& 0.0 & 0.0 \\
Argument of periastron, $\omega$ (rad)                          & - & $2.82_{-0.47}^{+0.58}$ & 0.0 & 0.0 & $\mathcal{U}(0, 2\pi)$ & $3.29_{-0.44}^{+0.41}$ & 0.0 & 0.0 \\
\textbf{\textit{Other parameters}}                              & & & & & & & & \\
RV semi-amplitude, $K$ ($m s^{-1}$)                            & - & $3.88_{-0.8} ^{+0.82}$ & - & - & - & $3.64_{-0.69}^{+0.68}$ & - & $2.98 \pm 0.64$ \\
$^{\dagger}$ HIRES offset, $\gamma_{HIRES}$ ($m s^{-1}$)     & - & -0.18 $\pm$ 0.66 & - & - & $\mathcal{N}(0.0, 1.0)$ & $-0.49_{-0.54}^{+0.55}$ & $\mathcal{N}(0.0, 1.0)$ & -0.52 $\pm$ 0.54 \\
$^{\dagger}$ APF offset, $\gamma_{APF}$ ($m s^{-1}$)         & -  & -0.39 $\pm$ 0.72 & - & - & $\mathcal{N}(0.0, 1.0)$ & 0.52 $\pm$ 0.57 & $\mathcal{N}(0.0, 1.0)$ & 0.43 $\pm$ 0.58 \\
$^{\dagger}$ Background acceleration, $\dot{\gamma}$ ($m s^{-1} d^{-1}$) & - & $-0.007^{+0.0029}_{-0.003}$ & - & - & $\mathcal{N}(0.0, 1.0)$ & $-0.0067 \pm 0.0023$ & $\mathcal{N}(0.0, 1.0)$ & -0.0069 $\pm$ 0.0023 \\
$^{\ddagger}$ TESS jitter, $\log s_{TESS}$                    & - & - & $\mathcal{N}(-7.45, 10)$ & $-8.968_{-0.081}^{+0.069}$ & $\mathcal{N}(-7.45, 10)$ & $-8.968_{-0.080}^{+0.069}$ & $\mathcal{N}(-7.45, 10)$ & $-8.969_{-0.081}^{+0.070}$ \\
$^{\ddagger}$ HIRES jitter, $\log{s_{HIRES}}$ ($m s^{-1}$)    & $\mathcal{U}(-20, 20)$ & $5.0^{+0.56}_{-0.47}$ & - & - & $\mathcal{N}(0.52, 5)$ & 1.60 $\pm$ 0.10 & $\mathcal{N}(0.52, 5)$ & 1.58 $\pm$ 0.10 \\
$^{\ddagger}$ APF jitter, $\log{s_{APF}}$ ($m s^{-1}$)        & $\mathcal{U}(-20, 20)$  & $5.5^{+0.77}_{-0.75}$ & - & - & $\mathcal{N}(1.74, 5)$ & $1.59_{-0.14}^{+0.13}$ & $\mathcal{N}(1.74, 5)$ & $1.67_{-0.13}^{+0.12}$ 
\enddata
\tablenotetext{$*$}{Median posterior values are shown with the 68$\%$ confidence interval.}
\tablenotetext{$**$}{The quadratic limb darkening coefficients are implemented using the \citet{kipping+2013MNRAS} parameterization.}
\tablenotetext{\dagger}{The RV trend coefficients are for a linear function: $v_{r}(t) = \dot{\gamma}t + \gamma$.}
\tablenotetext{\ddagger}{Each jitter term was added to the respective data uncertainties in quadrature, such that the effective error was $\sqrt{\sigma^2 + s^2}$.}
\tablecomments{$\mathcal{U}(a, b)$ is a uniform distribution that is nonzero only between $a$ and $b$; $\mathcal{N}(\mu, \sigma)$ is a normal distribution with mean $\mu$ and standard deviation $\sigma$; $\mathcal{T}(\mu, \sigma, a, b)$ is a truncated normal distribution with mean $\mu$ and standard deviation $\sigma$, bounded between $a$ and $b$; and $\mathcal{L}(\mu, \sigma)$ is a log-normal distribution with log mean $\mu$ and log standard deviation $\sigma$.}
\end{deluxetable*}
\end{longrotatetable}

TOI-1751~b has a longer orbital period ($37.47 \rm{d}$) than $\sim90\%$ of all confirmed planets with measured radii consistent to 1$\sigma$ with TOI-1751~b and with well-constrained masses ($\frac{M_{pl}}{\sigma_{M_{pl}}}>3$). Additionally, it has a larger mass than $94\%$ of this sample, (see Fig. \ref{fig:mr_diagram}). It is also one of only 5 sub-Neptunes with periods longer than 30 days orbiting bright (V $<$ 10) stars, representing one of the best cases for investigating the warm sub-Neptune population.

TOI-1751~b is in a similar region of mass-radius space to several confirmed planets: TOI-561~d \citep{lacedelli+21}, TOI-1052~b \citep{armstrong+23}, TOI-1260~c \citep{georgieva+21}, Kepler-48~c \citep{steffen+13}, Kepler-107~e \citep{rowe+14}, K2-199~c \citep{mayo+18}, and Kepler-276~c and~d \citep{xie2014}. The analyses of these other sub-Neptunes are hindered by degeneracies between volatile-rich and rocky interiors, which are also present for TOI-1751~b.

Although TOI-1751 is likely a thin-disk star, metal-poor Neptune-hosting stars are typically enriched in alpha-elements \citep{adibekyan+12}. This suggests that planets orbiting these stars had fewer metals, and more water, available when forming. We thus favour a volatile-rich interior composition for TOI-1751~b (see Section \ref{sec:int}).

\subsection{Interior Composition} \label{sec:int}
The number of exoplanets with radii greater than $\sim3 R_\oplus$ declines sharply, with planets between 2.7 and 3.0 $R_\oplus$ around FGK stars within 100 d being up to 10 times more common than planets between 3.3 and 3.7 $R_\oplus$ \citep{fulton+18, hsu+19}. Currently, the only proposed mechanism for this phenomenon is the ``fugacity crisis'': the increased solubility of hydrogen in magma at high pressures \citep{kite+19}. At $\sim3 R_\oplus$, the pressures at the base of a planetary atmosphere are sufficient to sequester atmospheric hydrogen in magma. The associated loss of $\rm{H_2}$ (and thus, volume) from the atmosphere during formation may thus decrease the occurrence of $\gtrsim 3 R_\oplus$ planets. This makes an interesting case for studying the interior composition of planets with a radius just below $\sim3 R_\oplus$ such as TOI-1751~b. These planets may have sequestered hydrogen from their atmospheres, which may be observable as low atmospheric mass fractions.

The radius of TOI-1751~b (\Rpl{}) implies that it is unlikely to be a purely rocky planet \citep{rogers+15}. Furthermore, the instellation ($\sim 75 \rm{S_\oplus}$) and escape velocity ($\sim 25 \rm{km/s}$) of TOI-1751~b place it firmly on the right of the cosmic shoreline, implying that this planet has an atmosphere \citep{shoreline}. 

\begin{figure*}[bth]
    \centering
    \includegraphics[width=\textwidth]{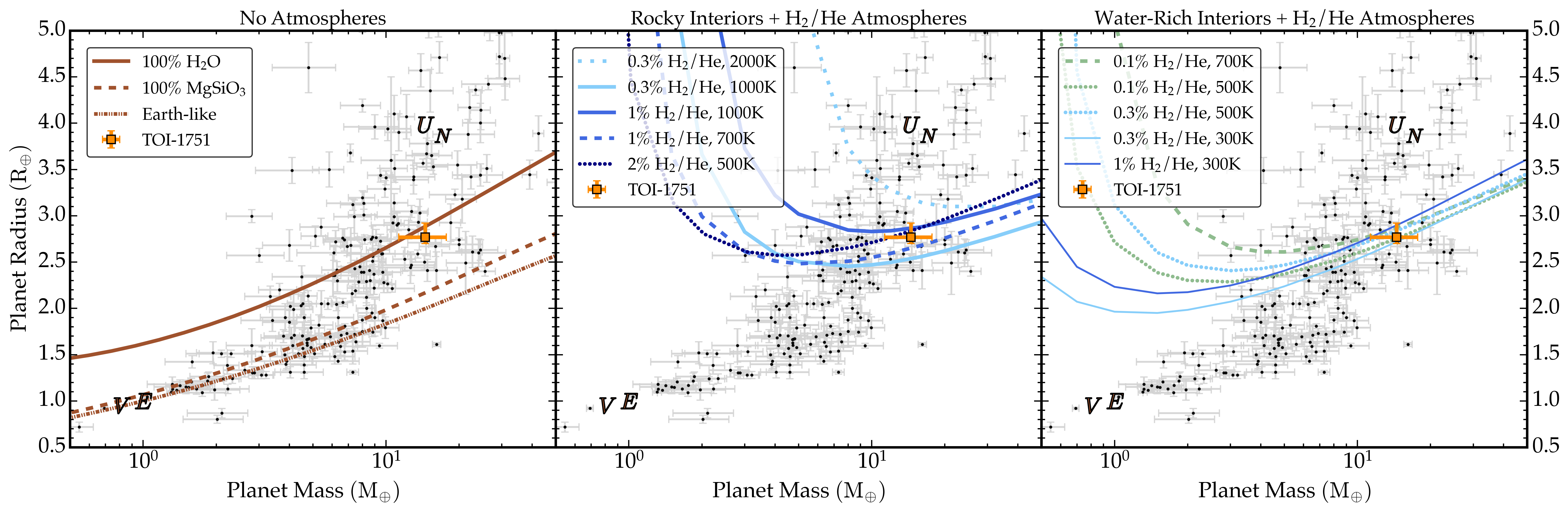}
    \caption{Mass vs. radius diagram showing confirmed planets with masses and radii measured to $>3\sigma$ precision (black points) and TOI-1751~b (orange). Each panel includes different theoretical composition curves from \citet{zeng+2019} . Earth (E), Venus (V), Uranus (U), and Neptune (N) are also shown for context, with the precise masses and radii for these planets lying in the center of the letter symbol. TOI-1751~b is consistent with both rocky and water-rich models. \textit{Left:} The Earth-like composition is assumed to be $32.5\%$ Fe/Ni-metal plus $67.5\%~\rm{MgSiO_3}$-rock. Curves for 100\% $\rm{H_2O}$, and $100\%~\rm{MgSiO_3}$ are also shown. \textit{Middle:} Composition curves assuming an Earth-like planet with the addition of a $\rm{H_2}$/He atmosphere (made up of a mixture of $75\%\ \rm{H_2}$ and $25\%$ He). These curves are evaluated along interior adiabats at different internal specific entropies, labelled by the temperature of the corresponding specific entropy at 100-bar level in the gas envelope (2000K is sparsely dashed, 1000K is solid, 700K is dashed, 500K is dotted, 300K is thin solid). \textit{Right:} Composition curves correspond to an Earth-like planet with an H/He isothermal envelope at various surface temperatures atop an ice-VII (a cubic crystalline form of ice) layer.}
    \label{fig:mr_diagram}
\end{figure*}

However, mass and radius measurements alone are not sufficient to uniquely constrain planet composition \citep{adams+08, rogers+10a}. The planet's bulk density is consistent with both a rocky interior with a few percent by mass H/He atmosphere and a volatile-rich interior with a H/He atmosphere (see Fig. \ref{fig:mr_diagram}). This degeneracy between silicate- and water-rich models is compounded by the unconstrained albedo and heat redistribution patterns, and is an ongoing challenge in characterizing sub-Neptunes \citep[e.g.][]{valencia+13, nixon+21, luque+22}. 

As shown in Fig. \ref{fig:mr_diagram}, if TOI-1751~b has a rocky interior, it will have a higher H/He atmospheric mass fraction compared to if it has a volatile-rich interior. A volatile-rich planet has a lower bulk density than that of a planet with an iron-rich core, and thus needs a less massive atmosphere to make up the bulk density. The models described in \citet{zeng+2019} are also consistent with those described in \citet{lopez+fortney2013}, which predict a 0.5-3$\%$ H/He atmosphere by mass for TOI-1751~b.

The relatively low atmospheric mass fraction of TOI-1751~b may have been sculpted by mass loss during evolution. Ongoing mass loss may be probed through future spectroscopic observations of Lyman $\alpha$ or meta-stable helium \citep[e.g.][]{spake+18, kulow+14}, although the planet's long transit duration will make these observations resource-intensive. Although this planet has a low Transmission Spectroscopy Metric \citep[TSM = 20,][]{kempton+18}, it may yet prove an interesting target for comparative atmospheric studies with next-generation instruments.

The mass, radius, and instellation of TOI-1751~b are also consistent with models of putative irradiated ocean worlds \citep{aguichine+2021}. These models are composed of refractory layers (iron core and rocky mantle), a hydrosphere with an equation of state that extends to the plasma regime, and a steam atmosphere. Within this framework, TOI-1751~b is consistent with a $\sim20-50\%$ water mass fraction for irradiation temperatures between 600K and 1000K ($T_{irr} = 700 K$ for TOI-1751~b, see Eqn. 9 of \citealt{aguichine+2021}). The instellation of TOI-1751~b at its current location also lies above the water vapour runaway greenhouse threshold \citep{shoreline}, suggesting that any water in this planet's atmosphere may exist as steam.

\begin{table*}[]
    \hspace*{-1cm}
    \begin{tabular}{|c|c|c|c|c|c|c|c|}
    \hline
    Family & CMF & MMF & WMF & AMF & Reference & $\rm{R_{p} (R_\oplus)}$ \\ 
    \hline
    \hline
    Earth-like & 0.325 & 0.675 & 0 & 0 & \citet{seager+07} & 1.97 \\
    Kepler-11c-like & 0.18 & 0.49 & 0.33 & 0 & \citet{acuna+22} & 2.67 \\
    Ganymede-like & 0.065 & 0.485 & 0.45 & 0 & \citet{seager+07} & 2.85 \\
    Neptune-like & 0.125 & 0.125 & 0.62 & 0.13 & \citet{podolak+95} & 4.99 \\
    Uranus-like & 0.02 & 0.02 & 0.92 & 0.04 & \citet{podolak+95} & 4.27 \\
    TOI-1751~b best fit & 0.36 & 0.53 & 0.10 & 0.01 & This work & 2.77 \\
    \hline
    \end{tabular}
    \caption{Compositional ``families'' of planet interiors that we used as inputs for the default \texttt{MAGRATHEA} mode to model TOI-1751~b in Section \ref{sec:default}. For each model we include the core mass fraction (CMF), mantle mass fraction (MMF), water mass fraction (WMF), and atmospheric mass fraction (AMF).}
    \label{tab:families}
\end{table*}

\begin{figure*}
    \centering
    \includegraphics[width=\textwidth]{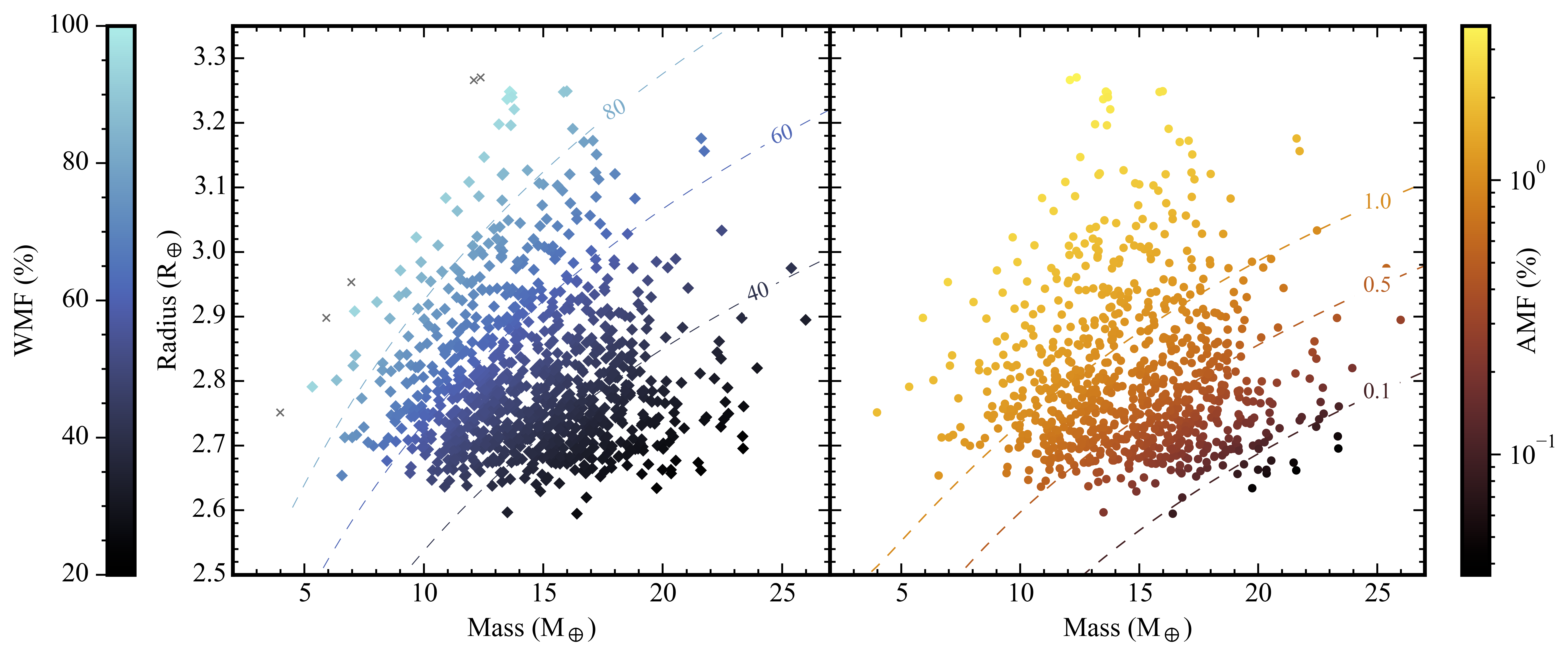}
    \caption{Mass vs. radius posterior samples from the joint transit and RV fit for TOI-1751~b modeled with \texttt{MAGRATHEA}. \textit{Left:} Models of TOI-1751~b assuming an Earth-like rock/iron ratio, no atmosphere (i.e., AMF = 0\%), a supercritical and condensed water layer, and a water envelope. By allowing the planet's envelope to be made of water, the WMF required to reproduce the planet's radius is substantially reduced. Grey x's mark mass and radius samples which are too low density to be modeled with AMF=0\%. \textit{Right:} Models of TOI-1751~b assuming an Earth-like mantle/core ratio, a WMF of 20\%, and a non-ideal H/He atmosphere. The non-ideal atmosphere is less compressible than an ideal H/He envelope, and thus a smaller AMF is needed to reproduce the measured planet radius. Iso-composition curves for WMF and AMF are shown with dashed lines.}
    \label{fig:wmfamf}
\end{figure*}

We used \texttt{MAGRATHEA}\footnote{\url{https://github.com/Huang-CL/Magrathea}} \citep{Huang2022} to investigate the composition of TOI-1751~b. \texttt{MAGRATHEA} is a 1D structure code which assumes fully differentiated planet layers: an iron core, a silicate mantle, a steam/water/ice hydrosphere, and a  H/He atmosphere. 

We note that these models do not encapsulate the true complexity of planet interiors, which may also include mixing between rock and water layers \citep{kovacevic+22, vazan+22}. Ice and rock may remain mixed in planetary interiors for billions of years if no significant mass loss occurs. The timescale for a planet to lose its atmosphere via XUV-driven photoevaporation scales with $M_p^2$ (or, more precisely, $M_{core}^2$) and scales inversely with instellation \citep[see Eqn. 4 in][]{lopez+fortney2013}. A massive sub-Neptune-sized planet such as TOI-1751~b may have been able to resist photoevaporative mass-loss and thereby retain a small but non-negligible H/He atmosphere, and thus, an ice-rock mixture, in its interior over its lifetime. 

We used two \texttt{MAGRATHEA} modes to model TOI-1751~b: the default model and a composition solver, which we discuss in Sections \ref{sec:default} and \ref{sec:solver}, respectively.

\subsubsection{Solving for Planet Radius} \label{sec:default}
Given the inherent degeneracies present in modelling sub-Neptune interiors, we consider several compositional  ``families'' of solutions based on Solar System objects and confirmed exoplanets in our analysis, differing in their core, mantle, water, and atmospheric mass fractions (CMF, MMF, WMF, and AMF, respectively). We summarize the families used in Table \ref{tab:families}.

The default \texttt{MAGRATHEA} mode calculates the planet radius for a given planet mass and distribution of mass between layers. For each step in enclosed mass, the pressure, density, and temperature are calculated, moving from $M=0$ to $M=M_p$. This mode takes as input the temperature discontinuities at each boundary layer (if any), as well as the surface temperature. We assume thermal equilibrium between each boundary layer and a surface temperature equal to the calculated equilibrium temperature (820 K). We use the default phase diagrams and equations of state for the iron core and magnesium silicate mantle. We use an updated version of \texttt{MAGRATHEA} which includes tabulated equations of state for water in liquid, solid, vapor, and supercritical forms in the hydrosphere (\citet{haldemann+20}) and for a H/He atmosphere with solar composition (\citet{Chabrier2021}).

We find that Uranus-like, Neptune-like, and Earth-like compositions do not reproduce the measured radius of TOI-1751~b. We find that a WMF between that of a Kepler-11c-like and Ganymede-like model most closely reproduced the measured radius of TOI-1751~b, though these objects are very different in scale and environment.

We also manually refined the mass distribution between layers in order to optimize the model. In this, we set an atmospheric mass fraction of 1\%, based on other models of planet composition \citep{zeng+2019, lopez+fortney2013}. The model which most closely reproduces the radius of TOI-1751~b includes a CMF of 36$\%$, MMF of 53$\%$, WMF of 10$\%$, and an AMF of 1$\%$. However, we stress that this is a possible composition rather than a prescriptive model of the interior of this planet.

\subsubsection{Solving for Mass Fractions} \label{sec:solver}
The second \texttt{MAGRATHEA} input method used, a composition solver, calculates a planet's atmospheric and water mass fractions given a fixed planet mass and radius. Again, we assumed thermal equilibrium between each boundary layer and calculated the null albedo surface temperature to be 690 K. We drew 1000 samples from the planet mass and radius posteriors measured using RV and photometric data, and used these as inputs for the composition solver. Our base model used an interior that has an Earth-like core-to-mantle ratio (32.5\% core to 67.5\% mantle). We then found the WMF (including a water-vapor atmosphere) required to match the radius of TOI-1751~b. As shown in Fig. \ref{fig:wmfamf}, we found WMFs from $21$ to $100\%$ with a median WMF of 49\%. This median WMF may suggest the planet formed beyond the snow line, much further out than its present location.

However, if we include a H/He atmosphere above supercritical water, the planet is consistent with a lower WMF. We now assumed an Earth-like core to mantle ratio and a WMF of 20\%, set by the lowest WMF found in the previous section. We then redistributed mass from the interior to form a H/He atmosphere. The atmosphere has an isothermal temperature profile with temperature increasing adiabatically at pressures above 100 bar (similar to the models of \citealt{nixon+21}). The AMF under this model ranges from 0.04-3.65\% with a median of 0.72\% (see Fig. \ref{fig:wmfamf}), much smaller than the atmospheric mass fractions of Neptune and Uranus. These findings are consistent with Section \ref{sec:default} and demonstrate the range of possible interior solutions for TOI-1751~b. 

\section{Conclusions} \label{sec:conc}
In this work, we study the solar-type star TOI-1751 using \tess{} photometry, and Keck/HIRES and APF/Levy RV observations. We confirm the planetary nature of TOI-1751~b, a transiting sub-Neptune, and measure its radius (\Rpl{}) and mass (\Mpl{}). From these measurements, we infer a bulk density of \rhopl{}. This points to several possible compositions: a rocky planet with a H/He atmosphere, a sub-Neptune with a volatile-rich interior, or an irradiated ocean planet with a volatile-rich interior and steam atmosphere. 

TOI-1751 is a metal-poor ([Fe/H] = $-0.40 \pm 0.06$) star, and as such, may have had a water-enriched protoplanetary disk \citep{adibekyan+12}. We thus favor a volatile-rich (i.e. metal-poor) interior for this planet. We find its bulk density (\rhopl{}) can be matched by a volatile-rich interior with an atmosphere mass fraction of $\sim1\%$. TOI-1751~b is a relatively long-period (37.47 d) but highly irradiated ($T_{eq} = 820K$) planet, giving us an insight into the small warm Neptune population with precisely measured masses and radii.

\facility{Keck:I (HIRES, \citealt{hires}), APF \citep{apf}, TESS \citep{tess}} 

\software{This research made use of \texttt{exoplanet} \citep{exoplanet:joss, exoplanet:zenodo} and its dependencies \citep{exoplanet:agol20, exoplanet:arviz, exoplanet:astropy13, exoplanet:astropy18, exoplanet:astropy22, exoplanet:pymc3, exoplanet:theano}. This research also made use of \texttt{Lightkurve}, a Python package for Kepler and TESS data analysis \citep{Lightkurve_2018}, and its dependencies \citep{exoplanet:astropy13, exoplanet:astropy18, astroquery, Astrocut}. Additionally, this work made use of the following Python packages: \texttt{numpy} \citep{numpy}, \texttt{pandas} \citep{reback2020pandas}, \texttt{matplotlib} \citep{matplotlib}, \texttt{corner} \citep{corner}, \texttt{RadVel} \citep{radvel}, \texttt{SpecMatch-Emp} \citep{specmatch-emp}, \texttt{SpecMatch-Synth} \citep{specmatch-synth}, and \texttt{CMasher} \citep{cmasher}. }

\begin{acknowledgments}
Some of the data presented in this paper were obtained at the W. M. Keck Observatory, which is operated as a scientific partnership among the California Institute of Technology, the University of California and the National Aeronautics and Space Administration. The Observatory was made possible by the generous financial support of the W. M. Keck Foundation. The authors wish to recognize and acknowledge the very significant cultural role and reverence that the summit of Maunakea has always had within the indigenous Hawaiian community. We are most fortunate to have the opportunity to conduct observations from this sacred mountain which is now colonized land.

This paper made use of data collected by the TESS mission and are publicly available from the Mikulski Archive for Space Telescopes (MAST) operated by the Space Telescope Science Institute (STScI). Funding for the TESS mission is provided by NASA’s Science Mission Directorate. We acknowledge the use of public TESS data from pipelines at the TESS Science Office and at the TESS Science Processing Operations Center. Resources supporting this work were provided by the NASA High-End Computing (HEC) Program through the NASA Advanced Supercomputing (NAS) Division at Ames Research Center for the production of the SPOC data products.

C. Brinkman, F. Dai, S. Giacalone, J. Lubin, J. Akana Murphy, M. Rice, and J. van Zandt contributed $>10$ observations of TOI-1751 using Keck/HIRES.

We thank Artem Aguichine for useful discussions on interpreting composition models, and Andreia Carrillo for insight on the galactic context of TOI-1751. We also acknowledge Erik Petigura and BJ Fulton for contributions to the construction, design, and team management of the \tess{}-Keck Survey.

M. Rice acknowledges support from Heising-Simons Foundation Grant. \#2023-4478. J. Akana Murphy is supported by the National Science Foundation (NSF) Graduate Research Fellowship Program (GRFP) under Grant No. DGE-1842400. A. Desai would like to acknowledge the UC Berkeley Physics Undergraduate Research Scholars Program, and the UC Berkeley Mathematical \& Physical Sciences Scholars for funding their work. E. Turtelboom acknowledges support from a David \& Lucile Packard Foundation grant. C. Harada acknowledges support from the National Science Foundation Graduate Research Fellowship Program under Grant No. DGE 2146752. D. Huber acknowledges support from the Alfred P. Sloan Foundation, the National Aeronautics and Space Administration (80NSSC21K0652), and the Australian Research Council (FT200100871). M. Hill would like to acknowledge NASA support via the FINESST Planetary Science Division, NASA award number 80NSSC21K1536.

\end{acknowledgments} 

\appendix
In this Appendix, we present the RV observations used in our analysis, as well as corner plots from the joint photometry and RV analysis.

\startlongtable
\begin{deluxetable*}{rrrrrr}
\tablecaption{TOI-1751 RV and S Value \label{tab:rvdata}}
\tablehead{\colhead{Time} & \colhead{RV} & \colhead{RV Error} & \colhead{S Value} & \colhead{S Value Error} & \colhead{Telescope/Instrument} \\ \colhead{(BJD-2457000)} & \colhead{(m/s)} & \colhead{(m/s)} & \colhead{} & \colhead{Error} & \colhead{}}
\startdata
1967.950094 &   5.017768 &  5.730939 &  0.122338 &       0.002 &     APF/Levy \\
1967.964422 &  -4.575683 &  5.527218 &  0.134032 &       0.002 &     APF/Levy \\
1967.978948 &  -2.429906 &  5.696315 &  0.133791 &       0.002 &     APF/Levy \\
2028.820286 &   2.559080 &  1.589164 &  0.121900 &       0.001 & Keck/HIRES \\
2046.821867 &   3.707988 &  2.554524 &  0.123700 &       0.001 & Keck/HIRES \\
2067.831578 &   0.041008 &  2.484040 &  0.124000 &       0.001 & Keck/HIRES \\
2088.734881 &   1.607211 &  1.535046 &  0.117800 &       0.001 & Keck/HIRES \\
2092.815251 &  -0.893631 &  1.529151 &  0.123400 &       0.001 & Keck/HIRES \\
2117.707073 &   5.236989 &  1.584269 &  0.124000 &       0.001 & Keck/HIRES \\
2120.713800 &   3.724947 &  1.564962 &  0.123600 &       0.001 & Keck/HIRES \\
2122.733467 &   6.318717 &  1.642816 &  0.126200 &       0.001 & Keck/HIRES \\
2153.700835 &   7.780182 &  1.990264 &  0.127000 &       0.001 & Keck/HIRES \\
2268.167963 &  -0.077914 &  1.746272 &  0.139400 &       0.001 & Keck/HIRES \\
2297.163692 &   4.810280 &  2.969652 &  0.129000 &       0.001 & Keck/HIRES \\
2299.986968 &  -4.555499 &  1.697425 &  0.137600 &       0.001 & Keck/HIRES \\
2301.948192 &  -5.943242 & 12.029358 &  0.133047 &       0.002 &     APF/Levy \\
2315.035716 &  29.038655 & 25.242176 & -0.083370 &       0.002 &     APF/Levy \\
2319.836634 & -15.225166 & 13.325574 &  0.076023 &       0.002 &     APF/Levy \\
2323.019259 &  10.611561 & 12.422581 &  0.099622 &       0.002 &     APF/Levy \\
2328.974946 & -19.033063 & 11.799784 &  0.127379 &       0.002 &     APF/Levy \\
2332.872891 & -32.112722 & 20.993940 &  0.122629 &       0.002 &     APF/Levy \\
2336.960193 & -21.378039 & 11.914451 &  0.134776 &       0.002 &     APF/Levy \\
2342.002709 & 134.689850 & 33.994404 &  0.196318 &       0.002 &     APF/Levy \\
2346.002758 &  -8.027158 & 14.547453 &  0.185963 &       0.002 &     APF/Levy \\
2358.890727 &  -1.986562 &  1.620713 &  0.138100 &       0.001 & Keck/HIRES \\
2367.964110 &   0.165681 &  2.162707 &  0.138000 &       0.001 & Keck/HIRES \\
2368.781398 &   1.358598 &  2.365940 &  0.139700 &       0.001 & Keck/HIRES \\
2377.061696 & -10.119392 &  1.456894 &  0.137800 &       0.001 & Keck/HIRES \\
2379.892405 &  -7.151634 &  1.572555 &  0.136400 &       0.001 & Keck/HIRES \\
2381.969148 &  -6.812479 &  1.651548 &  0.137300 &       0.001 & Keck/HIRES \\
2384.908952 &  -6.832657 &  1.628575 &  0.136500 &       0.001 & Keck/HIRES \\
2385.863678 &   1.057305 &  1.537402 &  0.137800 &       0.001 & Keck/HIRES \\
2387.881649 &  -8.906562 &  1.853298 &  0.138300 &       0.001 & Keck/HIRES \\
2389.029752 &  -3.754111 &  1.821554 &  0.138200 &       0.001 & Keck/HIRES \\
2395.889179 &  -3.276423 &  1.539202 &  0.137500 &       0.001 & Keck/HIRES \\
2399.877840 &   0.290966 &  1.622769 &  0.136800 &       0.001 & Keck/HIRES \\
2404.855951 &  -5.878335 &  6.827647 &  0.122974 &       0.002 &     APF/Levy \\
2404.983055 &  -0.196117 &  1.460241 &  0.137000 &       0.001 & Keck/HIRES \\
2407.884841 &  -0.949310 &  1.545025 &  0.000000 &       0.001 & Keck/HIRES \\
2411.786829 &  -8.196623 &  1.793193 &  0.137400 &       0.001 & Keck/HIRES \\
2412.959517 &  -9.350621 &  1.625912 &  0.138000 &       0.001 & Keck/HIRES \\
2414.003977 & -12.236013 &  8.510894 &  0.112257 &       0.002 &     APF/Levy \\
2414.968983 &  -7.469030 &  1.867751 &  0.139400 &       0.001 & Keck/HIRES \\
2419.921674 &  14.523581 &  6.515332 &  0.127170 &       0.002 &     APF/Levy \\
2420.843476 &  -2.169806 &  1.716373 &  0.137500 &       0.001 & Keck/HIRES \\
2427.886891 &  -4.590306 &  6.220817 &  0.123955 &       0.002 &     APF/Levy \\
2435.812873 &   2.603392 &  1.647762 &  0.137400 &       0.001 & Keck/HIRES \\
2441.881509 &  -3.739677 &  1.483798 &  0.138000 &       0.001 & Keck/HIRES \\
2446.892561 &  -3.117825 &  2.034989 &  0.142400 &       0.001 & Keck/HIRES \\
2450.815713 &  -1.924462 &  1.498762 &  0.138300 &       0.001 & Keck/HIRES \\
2452.805114 &   1.352993 &  1.438917 &  0.138000 &       0.001 & Keck/HIRES \\
2455.738040 &  -6.254118 &  1.423055 &  0.138100 &       0.001 & Keck/HIRES \\
2465.870699 & -10.648689 & 11.000022 &  0.158808 &       0.002 &     APF/Levy \\
2466.812847 &   4.092209 &  6.834493 &  0.123571 &       0.002 &     APF/Levy \\
2468.831819 &   8.989297 &  8.457207 &  0.119219 &       0.002 &     APF/Levy \\
2469.730819 &   2.537741 &  1.324689 &  0.137700 &       0.001 & Keck/HIRES \\
2469.799233 &  29.879020 &  5.730486 &  0.134830 &       0.002 &     APF/Levy \\
2470.725433 &  -2.017303 &  1.383176 &  0.137700 &       0.001 & Keck/HIRES \\
2470.842261 &  12.113155 &  7.543221 &  0.108196 &       0.002 &     APF/Levy \\
2471.717605 &   5.593577 &  1.397789 &  0.137400 &       0.001 & Keck/HIRES \\
2472.826884 &  12.578924 &  5.993051 &  0.137314 &       0.002 &     APF/Levy \\
2475.715672 &  -3.750316 &  1.575992 &  0.135200 &       0.001 & Keck/HIRES \\
2476.714570 &   2.513236 &  1.477444 &  0.138100 &       0.001 & Keck/HIRES \\
2489.748011 &  -2.775751 &  1.744298 &  0.137800 &       0.001 & Keck/HIRES \\
2497.710333 &   2.230133 &  2.140036 &  0.121900 &       0.001 & Keck/HIRES \\
2498.709760 &   5.294520 &  1.782478 &  0.137500 &       0.001 & Keck/HIRES \\
2502.764471 &   4.870154 &  1.705234 &  0.140300 &       0.001 & Keck/HIRES \\
2503.757997 &   5.214706 &  1.811255 &  0.139700 &       0.001 & Keck/HIRES \\
2506.701908 &   6.818792 &  1.851065 &  0.144200 &       0.001 & Keck/HIRES \\
2508.706566 &  12.245074 &  1.769580 &  0.142200 &       0.001 & Keck/HIRES \\
2513.710780 &   6.999444 &  1.776015 &  0.142100 &       0.001 & Keck/HIRES \\
2516.697559 &   8.130837 &  1.793256 &  0.143100 &       0.001 & Keck/HIRES \\
2527.060108 & -10.371655 &  8.321552 &  0.121344 &       0.002 &     APF/Levy \\
2542.023432 &   9.081054 &  7.249155 &  0.130074 &       0.002 &     APF/Levy \\
2544.024300 &  13.435542 &  6.287968 &  0.201886 &       0.002 &     APF/Levy \\
2548.004834 &   0.731266 &  5.575908 &  0.111315 &       0.002 &     APF/Levy \\
2551.001521 &   8.092444 &  6.440715 &  0.120030 &       0.002 &     APF/Levy \\
2553.582819 &   4.227784 &  6.748294 &  0.127320 &       0.002 &     APF/Levy \\
2590.888857 &   6.169268 & 10.346817 &  0.126366 &       0.002 &     APF/Levy \\
2592.056426 &   5.265382 &  5.732238 &  0.126658 &       0.002 &     APF/Levy \\
2593.169051 &   5.937546 &  1.878051 &  0.136700 &       0.001 & Keck/HIRES \\
2593.990826 &  -6.202685 &  5.860577 &  0.128780 &       0.002 &     APF/Levy \\
2595.051170 &  -5.758683 &  9.585746 &  0.274591 &       0.002 &     APF/Levy \\
2595.051170 &  -5.758683 &  9.585746 &  0.274591 &       0.002 &     APF/Levy \\
2596.067670 & -21.130907 & 28.493889 &       - &         - &     APF/Levy \\
2597.048695 &   1.958798 &  7.601473 &  0.131972 &       0.002 &     APF/Levy \\
2597.981189 &   3.550994 &  6.322966 &  0.135757 &       0.002 &     APF/Levy \\
2599.038639 &   1.837115 &  6.185833 &  0.134869 &       0.002 &     APF/Levy \\
2600.060115 &   8.172718 &  5.482867 &  0.135445 &       0.002 &     APF/Levy \\
2600.863140 &   8.045893 &  7.959724 &  0.143010 &       0.002 &     APF/Levy \\
2603.015851 &  -5.139369 &  5.140807 &  0.128947 &       0.002 &     APF/Levy \\
2604.941903 &  -5.993106 &  6.158518 &  0.140250 &       0.002 &     APF/Levy \\
2606.035683 &  18.474733 &  6.284544 &  0.145013 &       0.002 &     APF/Levy \\
2606.839204 &   0.440075 &  6.656321 &  0.133141 &       0.002 &     APF/Levy \\
2607.843364 &  -5.041286 &  7.796099 &  0.117363 &       0.002 &     APF/Levy \\
2610.023642 & -10.503228 &  6.140399 &  0.143567 &       0.002 &     APF/Levy \\
2611.014272 &  11.493749 &  7.911727 &  0.123162 &       0.002 &     APF/Levy \\
2617.894850 &   6.137322 &  5.560378 &  0.152855 &       0.002 &     APF/Levy \\
2618.886264 &   9.769615 &  5.670396 &  0.129258 &       0.002 &     APF/Levy \\
2622.079397 &  15.910851 &  2.009231 &  0.143300 &       0.001 & Keck/HIRES \\
2624.067432 &  -1.558083 &  5.490264 &  0.126063 &       0.002 &     APF/Levy \\
2625.043475 &   9.483644 &  5.689345 &  0.125615 &       0.002 &     APF/Levy \\
2626.115235 &  11.031161 &  1.960640 &  0.140300 &       0.001 & Keck/HIRES \\
2631.049307 &  -3.358169 &  5.790683 &  0.132574 &       0.002 &     APF/Levy \\
2632.058612 &   4.532732 &  1.858382 &  0.142900 &       0.001 & Keck/HIRES \\
2638.934775 &  -4.127176 &  6.093731 &  0.131248 &       0.002 &     APF/Levy \\
2639.914852 &  -4.024217 &  5.559092 &  0.119383 &       0.002 &     APF/Levy \\
2641.012419 &  -5.012459 &  5.453372 &  0.121262 &       0.002 &     APF/Levy \\
2642.042773 & -10.061118 &  8.734130 &  0.124421 &       0.002 &     APF/Levy \\
2647.006948 &   8.074206 &  5.063630 &  0.134188 &       0.002 &     APF/Levy \\
2648.876391 &  13.088721 &  8.273856 &  0.130011 &       0.002 &     APF/Levy \\
2649.934336 &  17.319730 &  6.780447 &  0.131638 &       0.002 &     APF/Levy \\
2655.005373 &   7.669171 &  6.624101 &  0.134112 &       0.002 &     APF/Levy \\
2655.124910 &   7.829334 &  1.786718 &  0.137800 &       0.001 & Keck/HIRES \\
2657.108873 &   5.187839 &  1.585076 &  0.138800 &       0.001 & Keck/HIRES \\
2659.924954 & -10.996333 &  6.670738 &  0.128516 &       0.002 &     APF/Levy \\
2661.051773 &   2.548310 &  1.743180 &  0.138900 &       0.001 & Keck/HIRES \\
2662.873567 &  -4.270035 &  5.088588 &  0.129136 &       0.002 &     APF/Levy \\
2663.873695 &  -0.381093 &  4.568761 &  0.126450 &       0.002 &     APF/Levy \\
2669.001235 &  11.278678 &  6.360282 &  0.134961 &       0.002 &     APF/Levy \\
2669.916859 &   4.973050 &  8.178942 &  0.123104 &       0.002 &     APF/Levy \\
2670.947898 &  -5.991317 &  5.077427 &  0.123113 &       0.002 &     APF/Levy \\
2671.866865 & -18.742595 &  5.446438 &  0.129167 &       0.002 &     APF/Levy \\
2672.005846 &  -0.682016 &  1.882599 &  0.138200 &       0.001 & Keck/HIRES \\
2674.890709 &  -4.694932 &  6.980327 &  0.139405 &       0.002 &     APF/Levy \\
2677.852501 & -24.448137 &  5.099940 &  0.134874 &       0.002 &     APF/Levy \\
2678.852936 & -10.454044 &  4.799206 &  0.128851 &       0.002 &     APF/Levy \\
2681.917363 &  -5.304394 &  1.559754 &  0.138600 &       0.001 & Keck/HIRES \\
2686.897698 &  -0.943478 &  5.158717 &  0.133215 &       0.002 &     APF/Levy \\
2687.856996 &  -5.210267 &  5.362441 &  0.123747 &       0.002 &     APF/Levy \\
2689.901771 &  -0.696835 &  6.252559 &  0.142686 &       0.002 &     APF/Levy \\
2693.861144 &  -1.362535 &  5.006232 &  0.115349 &       0.002 &     APF/Levy \\
2694.840289 &  -4.238488 &  5.493890 &  0.120752 &       0.002 &     APF/Levy \\
2695.885880 & -10.519235 &  8.561679 &  0.310315 &       0.002 &     APF/Levy \\
2696.835002 &  -4.040712 &  5.772600 &  0.119838 &       0.002 &     APF/Levy \\
2698.927259 &   4.005478 &  5.065654 &  0.118580 &       0.002 &     APF/Levy \\
2699.826832 &  -9.413585 &  5.445652 &  0.122652 &       0.002 &     APF/Levy \\
2700.860350 &   4.388364 &  6.088020 &  0.126920 &       0.002 &     APF/Levy \\
2700.979191 &  -0.262092 &  1.677178 &  0.139900 &       0.001 & Keck/HIRES \\
2701.818903 &   2.291902 &  5.248309 &  0.123027 &       0.002 &     APF/Levy \\
2703.828034 &   2.313856 &  5.804502 &  0.153793 &       0.002 &     APF/Levy \\
2704.843576 &   8.915670 &  4.829378 &  0.147600 &       0.002 &     APF/Levy \\
2705.900461 &   8.175372 &  4.735982 &  0.137716 &       0.002 &     APF/Levy \\
2711.829592 & -18.952053 &  6.325415 &  0.146889 &       0.002 &     APF/Levy \\
2711.953051 &  -3.213032 &  1.805574 &  0.138200 &       0.001 & Keck/HIRES \\
2713.844574 &  -8.904660 &  4.966397 &  0.121233 &       0.002 &     APF/Levy \\
2714.842557 &   2.580516 &  5.792560 &  0.127062 &       0.002 &     APF/Levy \\
2715.810378 & -13.409715 &  6.517595 &  0.115936 &       0.002 &     APF/Levy \\
2716.838961 & -12.732788 &  5.676785 &  0.128701 &       0.002 &     APF/Levy \\
2717.827684 &  -9.949889 &  5.413912 &  0.125470 &       0.002 &     APF/Levy \\
2718.803247 &  -7.529141 &  6.957226 &  0.149397 &       0.002 &     APF/Levy \\
2719.820361 &  -2.255389 &  5.416015 &  0.170752 &       0.002 &     APF/Levy \\
2720.803447 &  -8.207650 &  5.289296 &  0.141908 &       0.002 &     APF/Levy \\
2721.815282 &   1.041374 &  4.832710 &  0.132775 &       0.002 &     APF/Levy \\
2722.829918 &   0.200221 &  4.850303 &  0.129646 &       0.002 &     APF/Levy \\
2723.802703 &   0.621059 &  5.310606 &  0.134380 &       0.002 &     APF/Levy \\
2724.818773 &   1.259379 &  4.681048 &  0.133453 &       0.002 &     APF/Levy \\
2725.811905 &   2.620762 &  5.251647 &  0.115851 &       0.002 &     APF/Levy \\
2726.821135 &  -4.503616 &  6.320291 &  0.123298 &       0.002 &     APF/Levy \\
2729.806129 &   0.778730 &  5.536544 &  0.153661 &       0.002 &     APF/Levy \\
2731.787888 &   7.148658 &  5.123953 &  0.122040 &       0.002 &     APF/Levy \\
2732.786897 &   9.807237 &  6.138739 &  0.147769 &       0.002 &     APF/Levy \\
2733.978380 &  15.132094 &  5.537754 &  0.124916 &       0.002 &     APF/Levy \\
2737.796134 &   0.461368 &  5.656452 &  0.128683 &       0.002 &     APF/Levy \\
2738.837353 &  -7.696666 &  4.891762 &  0.121002 &       0.002 &     APF/Levy \\
2738.928428 &  -2.419228 &  1.671246 &  0.136800 &       0.001 & Keck/HIRES \\
2739.787289 &   1.067751 &  6.911901 &  0.127771 &       0.002 &     APF/Levy \\
2740.765268 &  -3.419261 &  6.976524 &  0.106144 &       0.002 &     APF/Levy \\
2740.797131 &  -1.732072 &  1.479232 &  0.137800 &       0.001 & Keck/HIRES \\
2741.773964 &   7.188090 &  6.173003 &  0.128661 &       0.002 &     APF/Levy \\
2742.787105 &  -6.602692 &  6.094021 &  0.115056 &       0.002 &     APF/Levy \\
2743.910301 & -13.453842 &  5.776292 &  0.126866 &       0.002 &     APF/Levy \\
2744.804045 &  -1.905972 &  4.996543 &  0.119682 &       0.002 &     APF/Levy \\
2745.782557 &   0.831002 &  5.226742 &  0.131543 &       0.002 &     APF/Levy \\
2746.809157 & -17.495799 &  5.603072 &  0.124575 &       0.002 &     APF/Levy \\
2750.822390 &  -3.359310 &  4.701025 &  0.119518 &       0.002 &     APF/Levy \\
2752.770245 &  -5.441301 &  5.274942 &  0.118118 &       0.002 &     APF/Levy \\
2753.773339 &  -3.221144 &  6.413606 &  0.131876 &       0.002 &     APF/Levy \\
2759.843638 &  -0.143953 &  4.845451 &  0.124923 &       0.002 &     APF/Levy \\
2760.859845 &  -2.784372 &  5.460371 &  0.111267 &       0.002 &     APF/Levy \\
2761.769746 &  16.073382 &  5.176810 &  0.133473 &       0.002 &     APF/Levy \\
2762.767597 &   9.436855 &  4.402338 &  0.118103 &       0.002 &     APF/Levy \\
2763.836061 &   4.822846 &  5.854585 &  0.124828 &       0.002 &     APF/Levy \\
2764.847928 &  -1.109111 &  5.330804 &  0.123972 &       0.002 &     APF/Levy \\
2765.981170 &  -6.380672 &  1.775496 &  0.138700 &       0.001 & Keck/HIRES \\
2767.929813 &  -8.560310 &  4.931138 &  0.123475 &       0.002 &     APF/Levy \\
2768.828211 &   7.918173 &  5.664408 &  0.129152 &       0.002 &     APF/Levy \\
2769.822775 &  13.143110 &  5.950137 &  0.153334 &       0.002 &     APF/Levy \\
2769.972333 &   0.372534 &  1.526587 &  0.137900 &       0.001 & Keck/HIRES \\
2770.765084 &   1.743069 &  6.075838 &  0.120430 &       0.002 &     APF/Levy \\
2771.754382 &   6.802229 &  5.701153 &  0.232347 &       0.002 &     APF/Levy \\
2772.825138 &  -2.008147 &  5.038103 &  0.113421 &       0.002 &     APF/Levy \\
2773.805120 &  -2.187094 &  4.883149 &  0.119301 &       0.002 &     APF/Levy \\
2774.865355 &  -6.831143 &  5.155893 &  0.119926 &       0.002 &     APF/Levy \\
2775.749494 &   2.609458 &  5.060140 &  0.129475 &       0.002 &     APF/Levy \\
2776.751975 &  -3.608514 &  4.885476 &  0.126344 &       0.002 &     APF/Levy \\
2778.752680 & -12.085360 &  5.208640 &  0.254129 &       0.002 &     APF/Levy \\
2779.779756 &  -2.924123 &  5.139664 &  0.129396 &       0.002 &     APF/Levy \\
2780.754774 &   5.197845 &  4.900618 &  0.127491 &       0.002 &     APF/Levy \\
2781.742742 &   6.478144 &  4.858816 &  0.123557 &       0.002 &     APF/Levy \\
2782.749253 & -10.695371 &  4.863077 &  0.135994 &       0.002 &     APF/Levy \\
2783.741897 &  -8.505215 &  4.956112 &  0.125596 &       0.002 &     APF/Levy \\
2785.741261 & -13.079080 &  5.663354 &  0.119900 &       0.002 &     APF/Levy \\
2787.772199 &  -7.535848 &  5.967904 &  0.124651 &       0.002 &     APF/Levy \\
2788.735308 &  -8.269857 &  6.430127 &  0.153984 &       0.002 &     APF/Levy \\
2789.746368 &  -2.395887 &  5.312854 &  0.129404 &       0.002 &     APF/Levy \\
2790.750727 &  -4.767741 &  6.143120 &  0.130641 &       0.002 &     APF/Levy \\
2800.820008 & -13.122384 &  2.634958 &  0.136700 &       0.001 & Keck/HIRES \\
2828.740113 &   1.537343 &  2.593053 &  0.142200 &       0.001 & Keck/HIRES \\
2972.084400 &   5.696147 &  4.716573 &  0.130909 &       0.002 &     APF/Levy \\
2972.912956 &   5.916581 &  6.856952 &  0.122151 &       0.002 &     APF/Levy \\
3040.082792 & -10.693290 &  2.534608 &  0.139400 &       0.001 & Keck/HIRES \\
3070.930568 &  -7.465276 &  2.530681 &  0.140700 &       0.001 & Keck/HIRES \\
3104.798844 &   1.547328 &  2.637302 &  0.140500 &       0.001 & Keck/HIRES \\
3138.958952 &  -7.741875 &  2.594161 &  0.140500 &       0.001 & Keck/HIRES \\
\enddata
\end{deluxetable*}

\begin{figure*}[bth]
    \centering
    \includegraphics[width=0.95\textwidth]{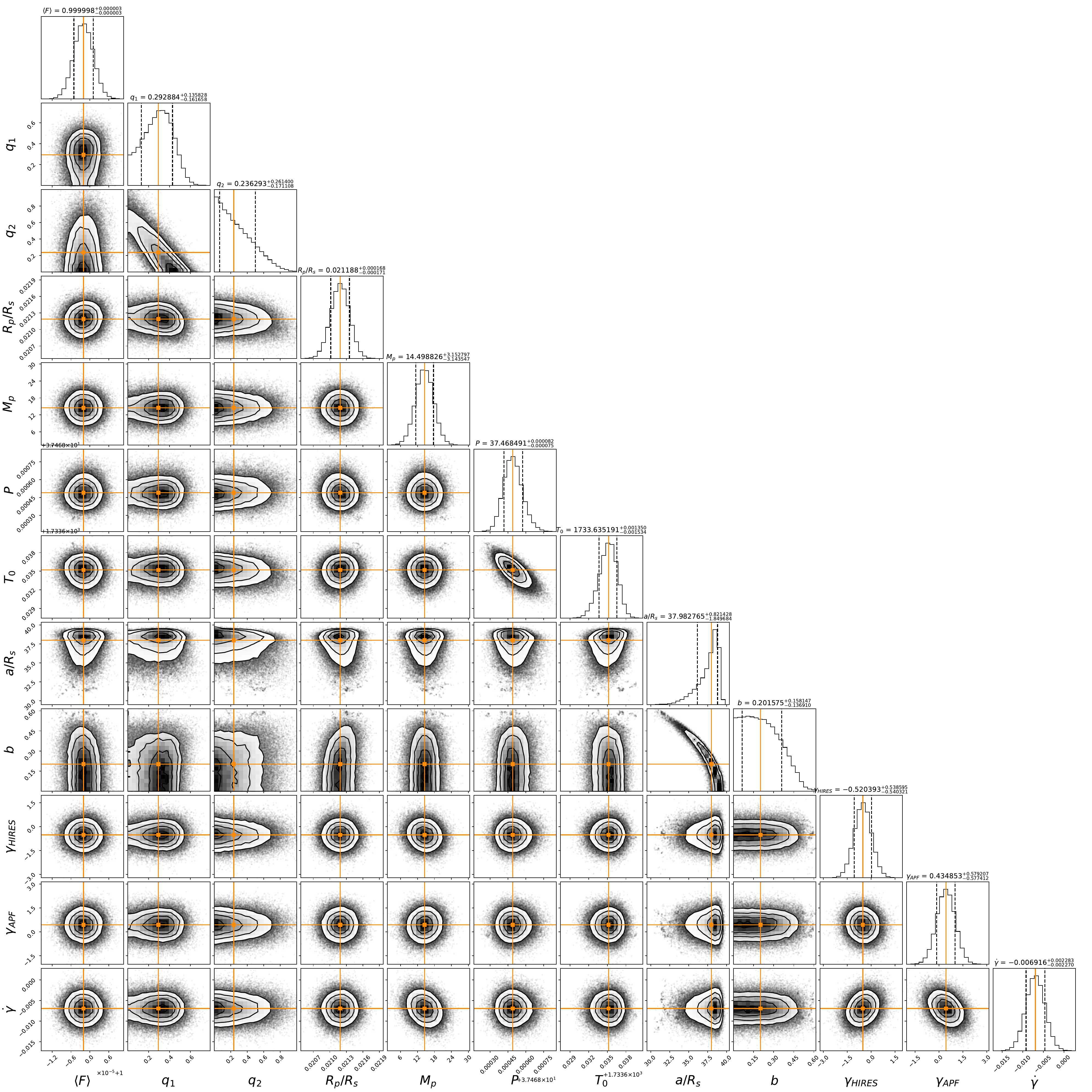}
    \caption{Posterior probability distributions for the joint RV and transit fit parameters for TOI-1751 b. The orange lines indicate the median value of the sample distribution, and the dashed black lines indicate 68\% interquantile range.}
    \label{fig:joint_b_posteriors}
\end{figure*}

\begin{figure*}[bth]
    \centering
    \includegraphics[width=0.95\textwidth]{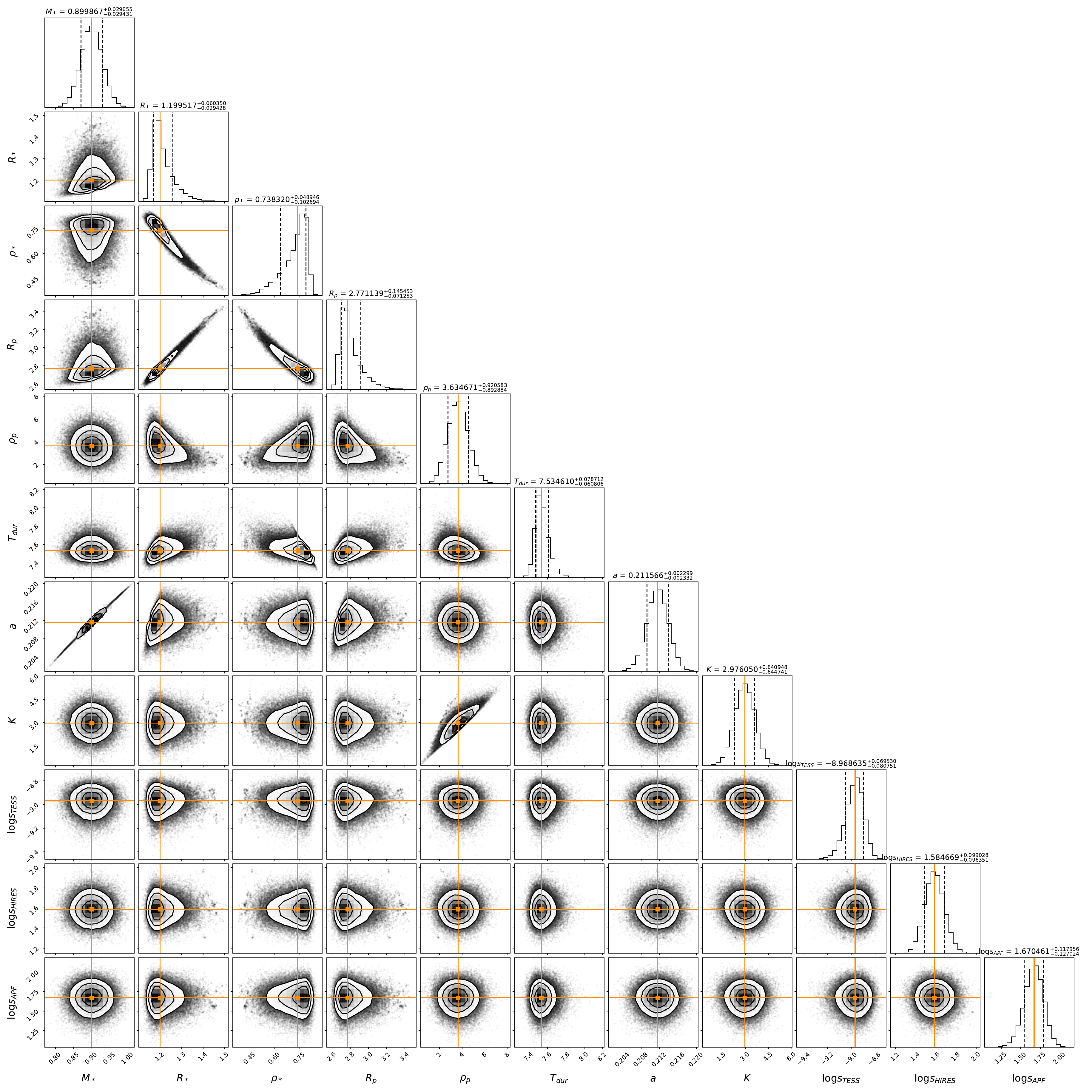}
    \caption{Auxiliary posterior distributions for the joint RV and transit fit parameters for TOI-1751 b. The orange lines indicate the median value of the sample distribution, and the dashed black lines indicate 68\% interquantile range.}
    \label{fig:joint_b_posteriors_aux}
\end{figure*}

\bibliography{main}{}
\bibliographystyle{aasjournal}

\end{document}